\documentclass[12pt,a4paper]{article}
\pdfoutput=1
\usepackage[latin1]{inputenc}
\usepackage{amsmath}
\usepackage{color}
\usepackage{amsfonts}
\usepackage{amssymb}
\usepackage{graphicx}
\usepackage{geometry}
\usepackage{amssymb,epsfig,subfigure}
\usepackage{amssymb}
\usepackage{hyperref}
\usepackage{comment}
\usepackage[font=footnotesize]{caption}
\usepackage[T1]{fontenc}
\usepackage{latexsym}
\usepackage[english]{babel}
\usepackage{cite}

\makeatletter
\renewcommand\section{\@startsection {section}{1}{\z@}%
                                 {-3.5ex \@plus -1ex \@minus -.2ex}
                                   {2.3ex \@plus.2ex}%
                                   {\normalfont\large\bfseries}}
\renewcommand\subsection{\@startsection{subsection}{2}{\z@}%
                                   {-3.25ex\@plus -1ex \@minus -.2ex}%
                                     {1.5ex \@plus .2ex}%
                                     {\normalfont\bfseries}}
\renewcommand\subsubsection{\@startsection{subsubsection}{3}{\z@}%
                                   {-3.25ex\@plus -1ex \@minus -.2ex}%
                                     {1.5ex \@plus .2ex}%
                                     {\normalfont\itshape}}
\makeatother

\def\pplogo{\vbox{\kern-\headheight\kern -29pt
\halign{##&##\hfil\cr&{\ppnumber}\cr\rule{0pt}{2.5ex}&\ppdate\cr}}}
\makeatletter
\def\ps@firstpage{\ps@empty \def\@oddhead{\hss\pplogo}%
  \let\@evenhead\@oddhead 
}
\def\maketitle{\par
 \begingroup
 \def\thefootnote{\fnsymbol{footnote}}
 \def\@makefnmark{\hbox{$^{\@thefnmark}$\hss}}
 \if@twocolumn
 \twocolumn[\@maketitle]
 \else \newpage
 \global\@topnum\z@ \@maketitle \fi\thispagestyle{firstpage}\@thanks
 \endgroup
 \setcounter{footnote}{0}
 \let\maketitle\relax
 \let\@maketitle\relax
 \gdef\@thanks{}\gdef\@author{}\gdef\@title{}\let\thanks\relax}
\makeatother

\numberwithin{equation}{section}

\newcommand\eea{\end{eqnarray}}
\newcommand\bea{\begin{eqnarray}}

\def\beq{\begin{equation}}
\def\eeq{\end{equation}}

\newcommand{\be}{\begin{equation}}
\newcommand{\ee}{\end{equation}}
\newcommand{\ba}{\begin{align}}
\newcommand{\ea}{\end{align}}
\newcommand{\bg}{\begin{gather}}
\newcommand{\eg}{\end{gather}}
\newcommand{\bseq}{\begin{subequations}}
\newcommand{\eseq}{\end{subequations}}

\begin{document}
\setcounter{page}0
\def\ppnumber{\vbox{\baselineskip14pt
}}
\def\ppdate{
} \date{}

\author{Horacio Casini\footnote{e-mail: casini@cab.cnea.gov.ar}, Marina Huerta\footnote{e-mail: marina.huerta@cab.cnea.gov.ar}, Javier M. Mag\'an\footnote{e-mail: javier.magan@cab.cnea.gov.ar}, Diego Pontello\footnote{e-mail: diego.pontello@ib.edu.ar} \\
[7mm] \\
{\normalsize \it Centro At\'omico Bariloche and CONICET}\\
{\normalsize \it S.C. de Bariloche, R\'io Negro, R8402AGP, Argentina}
}

\bigskip
\title{\bf  On the logarithmic coefficient of the entanglement entropy of a Maxwell field
\vskip 0.5cm}
\maketitle

\begin{abstract}
We elucidate the mismatch between the $A$-anomaly coefficient and the coefficient of the logarithmic term in the entanglement entropy of a Maxwell field.  In contrast to the usual assumptions about the protection of renormalization group charges at the infrared, the logarithmic term is different for a free Maxwell field and a Maxwell field interacting with heavy charges.   
This is possible because of the presence of superselection sectors in the IR theory. 
However, the correction due to the coupling with charged vacuum fluctuations, that restores the anomaly coefficient, is independent of the precise UV dynamics. The problem is invariant under electromagnetic duality, and the solution requires both the existence of electric charges and magnetic monopoles. We use a real-time operator approach but we also show how the results for the free and interacting fields are translated into an effective correction to the four-sphere partition function. 
\end{abstract}
\bigskip

\newpage

\tableofcontents

\vskip 1cm


\section{Introduction}

Entanglement entropy (EE) is an unconventional and useful theoretical quantity in the exploration of quantum field theories (QFT). It has been especially important in connection with holographic theories and the understanding of the renormalization group (RG) irreversibility. In extended quantum systems it has been a useful order parameter determining different types of quantum behavior.  It is always important in this line of research to establish a dictionary between entropic quantities and more conventional field-theoretic ones. An important and accepted entry in this dictionary is the identification of the coefficient of the logarithmic term in the EE for a conformal field theory in a sphere in even spacetime dimensions with the coefficient $A$ of the Euler term in the trace anomaly, \cite{Solodukhin:2008dh,Casini:2011kv,Herzog:2015ioa}
\be
S(R)= \cdots+ (-1)^{(d-2)/2} 4\, A \log(R/\delta)+\cdots\,, 
\ee     
with $R$ the radius of the sphere and $\delta$ a short distance cutoff. 

This identification follows from quite general and simple reasonings, and has been confirmed by direct computation for free scalars and fermion fields \cite{Casini:2010kt,Dowker:2010bu,Lohmayer:2009sq} as well as holographically \cite{Myers:2010xs}. However, it was noted by Dowker \cite{Dowker:2010bu} that a direct thermodynamic computation in de Sitter space for a free Maxwell field in $d=4$ fails to give the expected anomaly coefficient $-31/45$, giving instead a smaller coefficient $-16/45$, missing the anomaly by a correction of $-1/3$ (see an analogous calculation in hyperbolic space in \cite{Eling:2013aqa}). A confirmation of this conflictive result follows simply by decomposing the Maxwell field in spherical modes \cite{Casini:2015dsg}. There is a unitary mapping between the theory of two independent massless scalar fields and the one of a Maxwell field for all (decoupled) modes with angular momentum $l\ge 1$, and this unitary mapping is local in the radial coordinate. The $l=0$ mode is absent for the Maxwell field. This directly gives the logarithmic coefficient of the Maxwell field as $2\times (-1/90-1/6)=-16/45$, where $-1/90$ is the logarithmic coefficient for a scalar and $1/6$ is the one of the $l=0$ mode of the scalar, which corresponds to a one dimensional field with a positive logarithmic coefficient, whose entropy that has to be subtracted.       

This straightforward identification of operator algebras and states inside regions with spherical symmetry for the two theories leaves us no other alternative than to conclude that the anomaly does not match the logarithmic coefficient for the Maxwell field. We can also entertain the idea that the logarithmic term can be modified by the precise regularization procedure (or choice of algebra in a discretization of the theory). In that case, the same ambiguities would pollute the case of the scalar field modes for $l>0$, though the details of the choice of algebra for the two fields may be related to each other non locally along the boundary. 

 The questions that we address in this paper are what is special about the Maxwell field, why the proof of the identification with the anomaly goes wrong in this case, and under which circumstances the anomaly is recovered as the logarithmic coefficient of the EE. This last question is relevant to the entropic irreversibility theorem in $d=4$ \cite{Casini:2017vbe,Casini:2018kzx}.  

A possible solution was suggested in \cite{Casini:2015dsg} (see also \cite{Pretko:2018yxl}). There, it was speculated that while the pure Maxwell field has a specific coefficient that does not match the anomaly, this result might change in the presence of charged fields, which could, however, be very massive. The infrared (IR) theory is still free Maxwell. It contains superselection sectors for the different charges, and the constraints $\nabla E=\nabla B=0$ would be lifted by the charged fluctuations. 

While this might appear a natural proposal, it poses several important problems. The first one is that there are general arguments implying that a universal term like the one proportional to $\log(R)$ in the entropy for spheres of large radius is protected at the IR, i.e.,  it cannot be changed by changing the ultraviolet (UV) physics \cite{Grover:2011fa,Liu:2012eea}. These arguments are important for the assignation of this coefficient (for large spheres) to the physics of the IR fixed point of the theory. We address this question in the next section. A similar failure of the universal terms to be protected in the IR has been shown to happen in models with global superselection sectors \cite{Casini:2019kex}. 

A related problem is how a correction that depends on the details of the UV, such as the one associated with the presence of massive charges, could affect the IR result universally. This again is not restricted to the case of the Maxwell field but also happens for other models with superselection sectors \cite{Casini:2019kex}. The answer is that the main effect of charges is to destroy non-local correlations in some specific operators of the IR model. Hence, the result can be read off from the IR model itself irrespective of the precise UV physics. We will see how this happens in the Maxwell field in detail in section \ref{what}.          

In the literature, this problem is often blamed on the nature of gauge fields and solved in a way that does not subsist the continuum limit.
In fact, as we have already mentioned, this phenomenon is of much broader scope and does not have a direct relation with the description of the QFT in terms of gauge fields, which for some models may be a matter of choice,  but it occurs precisely when there are (gauge or global) superselection sectors in the infrared that are not present in the full theory. We will discuss in more detail the differences from our approach and previous works in the literature in section \ref{comparison}.

Most of the confusion around this subject comes from focusing on a bare entropy as the quantity of interest, which however does not have a clear physical meaning for the continuum model. The present problem is especially ill-posed in terms of the bare entropy. For example, in the context of our solution, one can ask how is it possible that a free model has a different coefficient that an interacting one independently of the size of the coupling constant. This discontinuity makes no sense unless one describes a physical quantity where the regulator $\epsilon$ is also physical. This can be done using the mutual information between two non intersecting regions $A$ and $B$, defined as 
\be
I(A,B)=\lim_{\delta\rightarrow 0} S_\delta(A)+S_\delta(B)-S_\delta(A\cup B)\,. \label{muu}
\ee
This limit for vanishing distance cutoff $\delta$ is finite and well defined. One can define a regularized entropy using the mutual information between a sphere of radius $R-\epsilon/2$ and the complement of a sphere of radius $R+\epsilon/2$ \cite{Casini:2015woa,Casini:2007dk}
\be
S_{\textrm{reg}}(R,\epsilon)=\frac{1}{2} I(R+\epsilon/2,R-\epsilon/2)\,.\label{muti}
\ee
The short distance $\epsilon$ is now physical. In these terms, our solution has the following form. For the pure Maxwell field we have 
\be
 S_{\textrm{reg}}\sim 4 \pi R^2\, \frac{k}{\epsilon^2}-\frac{16}{45} \log(R/\epsilon)+ \textrm{subleading}\,, 
\ee
with the ``incorrect'' logarithmic coefficient. The same result is expected for an interacting Maxwell field if $\epsilon$ is greater than the effective distance scale $\Lambda$ where the charge fluctuations become relevant, and which is set by the masses and couplings of the charged particles. In that case, the correlations between the two regions measured in the mutual information are the same as for the free field. We are in the IR regime and always keep $R\gg \Lambda, \epsilon$.  

Once we cross the scale of the charge fluctuations with our regulating distance, $\epsilon \ll \Lambda$, we have a modified result 
\be
 S_{\textrm{reg}}\sim 4 \pi R^2\, \left(\frac{k'}{\epsilon^2}- m^2\right)-\frac{31}{45} \log(R/\epsilon)+\textrm{subleading}\,, \label{ccc}
\ee
with the logarithmic term given by the anomaly, and where the missing terms are subleading in the large $R$ limit.  
The area term gets renormalized too, as expected, and the structure of the coefficient of the area term can have variations depending on the precise content of the UV theory. Here $m$ is a typical scale of the RG flow.   

Then, the question about the possible discontinuity of the logarithmic coefficient with the coupling constant has a natural explanation in terms of an order of limits. Whenever we make the coupling constant go to zero first than $\epsilon$, we get the free result, and the opposite limit gives us the anomaly. If we take the limit $\epsilon \rightarrow 0$ and $R\rightarrow \infty$ to define the logarithmic term in the IR (as required for the irreversibility theorems) we get two different results for interacting and exactly free fields, independently of the size of the interactions. 

One interesting and natural outcome of the calculation is that a full recovery of the anomaly coefficient requires magnetic monopoles along with electric charges. 

Finally, the last question is why the universal result for the interacting model numerically coincides with the anomaly. This question is addressed in section \ref{why}, where we discuss how to take into account the corrections for the free Maxwell field in the calculation based on the conformal mapping to de Sitter space.  
\section{How can massive charges correct the IR logarithmic coefficient? }
\label{physical}
Let us recall the argument for the protection of RG charges at the infrared \cite{Grover:2011fa,Liu:2012eea}. If we have the EE of a large region and change the UV physics keeping the IR theory invariant, the change will affect only correlations and entanglement at short distances across the boundary of the region. The change in one piece of the boundary is independent of the change in other pieces which are at an IR distance from it. Hence the result of this change in EE should be local and additive on the boundary. That is, it has the same general structure as the divergent terms of the EE. We expect it could be written for a general region as an integral over the boundary surface of local and geometric terms. The area term  can then be modified by the UV physics, but this is not the case of a $\log(R)$ term which cannot be produced by integrating a curvature tensor on the surface.\footnote{Note  the coefficient of $\log(\delta)$ can be changed and compensated by the logarithm of another dimensionful quantity.}  

This same argument can be translated in terms of the mutual information \cite{Casini:2015woa}. The question is now if for large $R$ the logarithmic term can be changed by changing $\epsilon$, where we are already in the regime $\epsilon \ll R$, or, equivalently, if it can be changed by altering the UV physics and keeping $\epsilon\ll R$ fixed.  We see from  (\ref{muu}) that as we change $\epsilon$ the change in the mutual information can only come from the entropy of the union of the two regions $S(A,B)$. This is equal to the entropy of the complement, that here is a thin spherical shell $r\in (R-\epsilon/2,R+\epsilon/2)$.\footnote{The identification of entropies of complementary regions is valid under the assumption of Haag duality, that is, that the algebra of the complementary region coincides with the commutant of the algebra of the region. This is valid for the Maxwell field in the present geometry of two nearly complementary balls, but this is not the case for theories with global superselection sectors. See \cite{Casini:2019kex}. }
 Then the argument is now that a thin shell should have an entropy that is local and additive along its surface \cite{Casini:2015woa}. This would guarantee the locality of the possible changes with $\epsilon$ and the UV physics, and the protection of the RG charges. 
 
 \begin{figure}[t]
\begin{center}  
\includegraphics[width=0.55\textwidth]{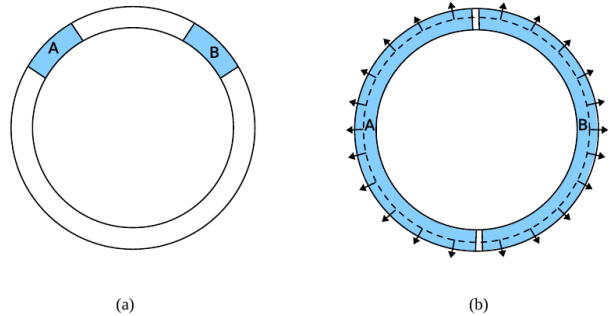}
\captionsetup{width=0.9\textwidth}
\caption{(a) Two regions of fixed angular span in the shell have vanishing mutual information in the $\epsilon\rightarrow 0$ limit. (b) Global constraints such as the vanishing of the electric flux may lead to non trivial correlations across the shell. }
\label{f2}
\end{center}  
\end{figure}

  Indeed, there is a simple reason why extensivity is expected as natural property for thin shells.  Extensivity can be partially rephrased as that mutual information between different parts of the shell vanishes in the limit of small width. This is because mutual information measures exactly the degree of non-extensivity of the entropy. But mutual information between two patches of the shell separated by a fixed distance should tend to zero in the limit of zero width for any theory (see figure \ref{f2}(a)). This is because the algebras of these shell patches do not contain any operator in the limit of $\epsilon\rightarrow 0$, and the correlations are kept bounded as we take the limit. There are no bounded operators that can be localized in a $d-2$ dimensional patch in QFT. In other words, when an operator becomes very thin it will be much more correlated with itself than with any other distant operator.

 \begin{figure}[t]
\begin{center}  
\includegraphics[width=0.45\textwidth]{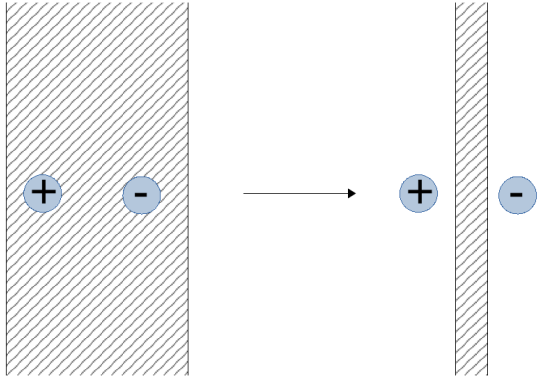}
\captionsetup{width=0.9\textwidth}
\caption{For $\epsilon$ wide enough and a smooth smearing of the electric flux charge anticharge fluctuations are averaged to zero and do not affect the fluctuations of the electric flux. When the width $\epsilon$ becomes small to allow for charge anticharge fluctuations to occur on each side of the wall, fluctuations of the electric flux will be large. }
\label{fig6}
\end{center}
\end{figure}

Given that, we can still identify a possible origin for the violation of extensivity. 
For this, we consider the case of two patches separated by a small distance $\epsilon$ of the order of the shell width, see figure \ref{f2}(b). In this scenario we cannot use the same reasoning. We do not consider these patches touching each other since we are not interested in UV divergent pieces of this mutual information but in the building up of long-distance correlations. These can appear because of the presence of constraints. For example, for the Maxwell field, the electric (or magnetic) fluxes $\Phi_1$, $\Phi_2$, over the two half-shells are not constrained, while the flux over the full shell has to vanish in absence of electric or magnetic charges, $\Phi_1+\Phi_2=0$. This implies correlations across the shell that are long-distance and non-extensive. Similar charge measuring operators appear in topological models, and more generally, in all models containing superselection sectors. Mutual information between nearby patches on the shell will notice these correlations. 

Hence, we have some non-extensivity of the shell entropy related to constraints. These constraints are modified when charges are added to the model and we are in a situation where charge fluctuations become important. This gives a physical explanation of the origin of the change in extensivity of entropy of the shell (and the change of the RG charges) when there is a transition from  $\epsilon\gg \Lambda$ to $\epsilon\ll \Lambda$, with $\Lambda$ a distance scale where the charge fluctuations affect the flux operators in the shell. 

More concretely, the Gauss law for the electric flux operator for the Maxwell field produces significant correlations on the shell. For the pure Maxwell field, we have for example $e^{i \Phi}=1$ for the total flux $\Phi$ across the shell. But how does this change when there are charges? The electric flux has to be smeared to become an operator in the shell algebra. When the width $\sim \epsilon$ of the operator smearing is much larger than the typical size separating the charge-anti-charge fluctuations in vacuum, these fluctuations will be averaged on the zone were the electric flux operator changes smoothly, and then the total flux will be zero as in the model without charges. See figure \ref{fig6}. We would have $\langle e^{i \Phi}\rangle =1$.  However, in the limit of small $\epsilon$, the charge fluctuations on each side of the shell will introduce large fluctuations to the flux operator seated on the shell. The expectation value $\langle e^{i \Phi}\rangle \simeq 0$ will vanish eliminating the long-distance correlations in the shell. In the presence of charges, the constraints become effective only for wide enough flux operators.

Let us see this more quantitatively. We can compute the vacuum fluctuations $\left\langle \Phi_{\Sigma}^{2}\right\rangle $ of the electric flux $\Phi_{\Sigma}=\int_{\Sigma}\bar{E}\left(\bar{x}\right)\cdot d\bar{S}$ across a patch $\Sigma$  on the shell. The correlation function of the electric field at equal time is
\begin{equation}
\left\langle E_{j}\left(0,\bar{x}\right)E_{k}\left(0\right)\right\rangle =\frac{1}{\left(2\pi\right)^{2}}\left(\partial_{j}\partial_{k}-\delta_{jk}\nabla^{2}\right)\frac{1}{\left|\bar{x}\right|^{2}}\,.\label{fefc}
\end{equation}
We should smear the flux of the radial electric field inside a thin shell of width  $\epsilon$ and compute the expectation value of the square of this operator.  Instead of smearing the electric field along the radial direction, a simpler calculation that shows the same essential features is to regularize the correlator changing $\left|\bar{x}\right|^{2}\rightarrow\left|\bar{x}\right|^{2}+\epsilon^{2}$ in \eqref{fefc}, such that the regularized correlator is still divergenless. We get
\be
\left\langle \Phi_{\Sigma}^{2}\right\rangle  = \frac{L_{\partial \Sigma}}{4\pi\epsilon}+\cdots\,,
\ee
where $L_{\partial \Sigma}$ is the perimeter of the boundary of the surface $\Sigma$. Therefore, the fluctuations satisfy a perimeter law. Indeed, the result can only depend on the perimeter since the normal fluxes across different surfaces sharing the same boundary are the same operators. This result is very peculiar of the conserved flux. It is not difficult to see that the fluctuations of other operators formed by an integral of a local field on the surface will have an area law.  This reduction in correlations to a perimeter law is clearly a consequence of Gauss law. 

Now, let us see what happens when the electromagnetic field is coupled with electric charges. In this case, we express the electric field Wightman correlator using its Kallen-Lehmann representation
\beq
\left\langle E_{j}\left(x\right)E_{k}\left(y\right)\right\rangle =\int_{0}^{+\infty}dq^{2}\rho\left(q^{2}\right)\,\int_{\mathbb{R}^{4}}\frac{d^{4}p}{\left(2\pi\right)^{4}}2\pi\,\Theta\left(p^{0}\right)\delta\left(p^{2}-q^{2}\right)\left[p_{0}^{2}\, \delta_{jk}-p_{j}p_{k}\right]\mathrm{e}^{-ip\cdot\left(x-y\right)}\,.\label{tiro}
\eeq
The spectral density function for the fields is, to lowest order in QED perturbation theory \cite{Itzykson:1980rh}, 
\beq
\rho\left(q^{2}\right)=Z \, \delta\left(q^{2}\right)+\frac{\alpha}{3\pi}\frac{1}{q^{2}}\left(1-\frac{4m_{e}^{2}}{q^{2}}\right)^{\frac{1}{2}}\left(1+\frac{2m_{e}^{2}}{q^{2}}\right)\Theta\left(q^{2}-4m_{e}^{2}\right)+\mathcal{O}\left(\alpha^{2}\right)\,,\label{spec}
\eeq
where $\alpha=\frac{e^{2}}{4\pi}$ is the fine-structure constant, $m_{e}$ is the electron mass, and $Z$ is the field renormalization constant. The first term with the delta function leads to the free field result with a divergenceless correlator (\ref{tiro}). The second term will give a different leading contribution to the flux fluctuations in the limit of small $\epsilon$, i.e. proportional to the area $A_\Sigma/\epsilon^2$ instead of the perimeter $L_{\partial \Sigma}/\epsilon$. 

To compute the coefficient of the area term, we compute the vacuum fluctuations of the total flux of the electric field on a planar surface, or more precisely
\beq
\Phi_{\Sigma}=\int d^{4}x\,E_{1}\left(x\right)f\left(x\right)\,,
\eeq
where the smearing function is  $f\left(x\right)=f_{0}\left(x^{0}\right)f_{1}\left(x^{1}\right)$.
The support of the smearing functions in $x^0$ and $x^1$ are restricted to the interval 
 $\left(-\frac{\epsilon}{2},\frac{\epsilon}{2}\right)$ 
 and we normalize $\int_{-\infty}^{+\infty}dx^{0}\,f_{0}\left(x\right)$ $= \int_{-\infty}^{+\infty}dx^{1}\,f_{1}\left(x^{1}\right)=1$.
Then, the vacuum fluctuations of the flux for a large patch of area $A_\Sigma$ in the plane is 
\begin{eqnarray}
&&\left\langle \Phi_{\Sigma}^{2}\right\rangle =\int d^{4}x\,d^{4}y\left\langle E_{1}\left(x\right)E_{1}\left(y\right)\right\rangle f\left(x\right)f\left(y\right)\nonumber\\&& 
\simeq A_\Sigma\, \int dx^0\,dx^1\,d^{4}y\left\langle E_{1}\left(x^0,x^1,0,0\right)E_{1}\left(y\right)\right\rangle f_0\left(x^0\right) f_0(y^0) f_1(x^1)f_1(y^1)\,\nonumber
\\
&&=\frac{A_\Sigma}{(2\pi)^2}\int d^2p\, \int_0^\infty dq^2\, q^2\,\rho(q^2)\, \theta(p^0)\, \delta(p^2-q^2)\, |\tilde{f}_0(p^0)|^2\, |\tilde{f}_1(p^1)|^2\,.\label{efi}
\end{eqnarray}
In the second line we have neglected a perimeter term. 

Since $q^2\rho(q^2) $ has support for $q^2\ge 4 m_e^2$, when the smearing functions are wide and smooth enough (and then $\epsilon m_e$ is large), their Fourier transform will be concentrated for small momentum and the integral will vanish exponentially in $\epsilon$. 

In the opposite limit of small $\epsilon m_e$ we roughly get  
\be
\left\langle \Phi_{\Sigma}^{2}\right\rangle \sim \left(\int_0^{\epsilon^{-2}}  dq^2\, \rho(q^2)\, q^2\right)  A_\Sigma\,,\label{en}
\ee
which by unitarity ($\rho(q^2)>0$) has a positive non zero coefficient. 
In the small distance limit the correlation of the charge density operators, which follows by taking divergences of (\ref{tiro}), is\footnote{The relation between charge density correlations and fluctuations of the electric flux on the surface follows directly from Gauss law. The flux over the surface is the charge on any side of it. Then, the self-correlation is equal to the correlation between total charges on each side of the surface.}
\be
\langle j^0(0) j^0(x)\rangle=\langle \nabla \cdot E(0)\, \nabla\cdot E(x)\rangle \sim  \frac{\left( \int_0^{x^{-2}}  dq^2\, \rho(q^2)\, q^2\right)}{x^4}\,.\label{fl}
\ee
If a scaling limit is reached for the current in the UV and the correlator of charge densities goes as $x^{-2 \Delta}$, then from the positivity of $\rho$ in (\ref{fl}) we must have $\Delta>2$ (see also \cite{Dorigoni:2009ra}). For a primary current in a CFT, $\Delta=3$, which given the asymptotic behavior of (\ref{spec}), corresponds to the case of the QED to this order.  Then, we have generically an area term in (\ref{en}) that is divergent with $\epsilon$ in the limit $\epsilon\rightarrow 0$.

A concrete result can be obtained for example using Gaussian smear functions
\beq
f_{0}\left(x\right)=f_{1}\left(x\right)=\frac{1}{\sqrt{\pi}\epsilon}\mathrm{e}^{-\frac{x^{2}}{\epsilon^{2}}}\,,
\eeq 
that are essentially localized in a size $\epsilon$. An straightforward computation gives for \eqref{efi} to the first order in $\alpha$ in QED
\beq
\left\langle \Phi_{\Sigma}^{2}\right\rangle =\alpha\,\,  g(m_e \, \epsilon)  \,\frac{A_\Sigma}{\epsilon^2}+{\cal O} \left(L_{\partial}/\epsilon\right)\,, \label{piry}
\eeq
where 
 the dimensionless function $g$ has an uninformative expression in terms of Meijer functions. The limits of the coefficient of the area term are 
\begin{eqnarray*}
m_{e}\epsilon\ll1: &  & g(m_e \, \epsilon)\sim \left(48\sqrt{2}\pi\Gamma\left(\frac{3}{4}\right)\Gamma\left(\frac{5}{4}\right)\right)^{-1}\,,\\
m_{e}\epsilon\gg1: &  & g(m_e \, \epsilon)\sim\frac{1}{8\sqrt{2}\left(2\pi\right)^{2}}\frac{\mathrm{e}^{-2\left(m_{e}\epsilon\right)^{2}}}{(m_{e}\epsilon)^{2}}\,.
\end{eqnarray*}
It is interesting that the turn on of the area term happens at a distance $\epsilon\sim m_e^{-1}$, independently  of the value of $\alpha$,  since it is given at this perturbative order by the statistics of charge fluctuations of free electrons.

In conclusion, we have a rather sharp transition between a perimeter law $L_{\partial \Sigma}/\epsilon$  for the fluctuations of the electric flux for large $\epsilon m_e$ (the limit of the pure Maxwell field) and an area law $\sim \alpha\, A_{\Sigma}/\epsilon^2$  for $\epsilon m_e$ small.  This transition gives a UV condition on the width of the smeared flux operator.  However, to have a transition in the flux fluctuations we need also an IR condition on the size of the flux operators,  
\be
R \gg \frac{\epsilon}{\alpha}\,,\label{yestaquetal}
\ee
such that the area term dominates over the perimeter one in (\ref{piry}). In the IR limit, this is always the case unless there are no interactions. In this sense, the qualitative change in the flux behavior is a non-perturbative effect that subsists for small $\alpha$.  

To show how this change in expectation values should lead to a change in the extensivity of the entropy,  we can take fluxes $\Phi_1$, $\Phi_2$, on two nearby patches on the shell, separated by a distance $\epsilon$ of the same order as the with of the shell. For Gaussian variables, the mutual information between the Abelian algebras generated by these operators is given by
\be
I=\frac{1}{2} \log \langle \Phi_1^2\rangle + \frac{1}{2} \log \langle \Phi_2^2\rangle-\frac{1}{2} \textrm{tr} \log \left(\begin{array}{cc}
\langle \Phi_1^2\rangle &\langle \Phi_1 \Phi_2\rangle \\
 \langle \Phi_1 \Phi_2\rangle & \langle \Phi_2^2\rangle
\end{array}\right)\,.  
\ee
For the free case, when the perimeters of the two regions are equal $L_1=L_2=L$, and the shared perimeter is $L_{12}$, we get
\be
I=\frac{1}{2}\log \left(\frac{L^2}{L^2-L_{12}^2}\right)\,.\label{fff}
\ee
This is independent of $\epsilon$ and shows there are important correlations along the surface that persist for any $\epsilon$ as long as the theory does not have charges. For the case of dominance of the area law, the flux operators are still effectively Gaussian variables since the fluctuations of the flux are produced by a large number of random independent charge fluctuations near the surface and we can apply the central limit theorem (see \cite{Casini:2019kex}). Since the areas of the nearby patches just add and $\langle \Phi_1 \Phi_2\rangle$ is still given by $\sim L_{12}/\epsilon$ we get 
\be 
I\sim \frac{\epsilon^2 \, L_{12}^2}{\alpha^2 A^2}\,,\label{iii}
\ee
where $A$ is the area of the patches. This is vanishing small if we have (\ref{yestaquetal}). 

The reason for this change is the large fluctuations acquired by each of the two flux operators while the correlation between them does not appreciably change. The main change is the elimination of the surprisingly large mutual information for the free Maxwell field (\ref{fff}) rather than the actual value of the small one of the interacting field (\ref{iii}). For small enough $\epsilon$ the difference just coincides with the free result (\ref{fff}) independently of the coupling $\alpha$. 
 Hence, this gives us the physical reason to expect universality for the correction, independently of the details of the interactions,  because the change is not due to some peculiar effect of the charges but rather resides in the destruction of a peculiar correlation present only for the free field.  
 If the are magnetic monopoles the same effect takes place for the fluxes of the magnetic field.   
 To find the form on which this change in extensivity affects the mutual information between the ball and its complement we need to take into account the full quantum algebra of the operators in the shell containing all flux operators in different patches at the same time. This is better done in an expansion of the operators in spherical variables as we do in the next section.   

\section{Calculation of the universal value of the correction}
\label{what}
Now we describe how the physical effect of heavy charges on the flux statistics across large surfaces described in the previous section is responsible for the change of the logarithmic coefficient of the entropy of a sphere. The prescription is clear and precise, we have to compute the mutual information for $R \gg m_e^{-1}, \alpha^{-1}\epsilon$, and evaluate the change of the coefficient of $\log(R)$ as we move $\epsilon  m_e$ from large to small values. Equivalently, we can evaluate the change for $R, \epsilon, m_e$ fixed, $R\gg m_e^{-1}\gg  \epsilon$, as we turn on the interactions. The exact computation can be quite difficult in an interacting theory. However, this should not be an obstacle to isolate and understand the contribution that produces the change in the logarithmic term since we are expecting a universal behavior in these two limits. For simplicity, we will think in terms of QED to lowest order in perturbation theory but, as it will become apparent in the following, the change in the logarithmic term does not depend on the details of the charged sector. 

The technical details of the calculation, as well as the final effective result, are in some aspects similar to the ones presented by Donnelly and Wall \cite{Donnelly:2015hxa,Donnelly:2014fua}, Soni and Trivedi \cite{Soni:2016ogt} (see also \cite{Moitra:2018lxn}), and Huang \cite{Huang:2014pfa} to solve the same problem. 
However, there are several important conceptual and quantitative differences, our calculation is very different in spirit from these works. We will discuss previous results in the literature in comparison with the present paper in section \ref{comparison}. We start by reviewing the case of the free Maxwell field in more detail. 

\subsection{Logarithmic coefficient for the free Maxwell field}
\label{frito}

Let us briefly review the case of the free Maxwell field on the sphere. See \cite{Casini:2015dsg} for a detailed discussion. This is the theory of electric and magnetic fields with equal time commutation relations
\be
[E^i(\vec{x}),B^j(\vec{y})]=i \varepsilon^{ijk}\, \partial_k \delta^3(\vec{x}-\vec{y})\,,\label{comu}
\ee
constraint equations 
\be
\vec{\nabla} \cdot \vec{E}=\vec{\nabla} \cdot \vec{B}=0\,,\label{cont}
\ee
and Hamiltonian
\be
H=\int d^3x\, \frac{1}{2}\left(\vec{E}^2+\vec{B}^2\right)\,.\label{hami}
\ee
The fields are Gaussian with two point correlators given by 
\bea
&&\left\langle E_{j}\left(0,\bar{x}\right)E_{k}\left(0\right)\right\rangle =\left\langle B_{j}\left(0,\bar{x}\right)B_{k}\left(0\right)\right\rangle=\\ &&\hspace{4cm}\left(2\pi\right)^{-2}\left(\partial_{j}\partial_{k}-\delta_{jk}\nabla^{2}\right)\frac{1}{\left|\bar{x}\right|^{2}}=\frac{1}{\pi^2} \left(2\frac{x_j x_k }{|\bar{x}|^6}-\frac{\delta_{ij}}{|\bar{x}|^4}\right)\,.\nonumber
\eea

 Taking into account the spherical symmetry of the problem, we expand the electric and magnetic fields in vector spherical harmonics 
\bea
\vec{E} &=& \sum_{l,m} E_{lm}^r(r,t) \vec{Y}_{lm}^r(\theta,\phi)+E_{lm}^e(r,t) \vec{Y}_{lm}^e(\theta,\phi)+E_{lm}^m(r,t) \vec{Y}^m_{lm}(\theta,\phi)\,,\label{18}\\
\vec{B} &=& \sum_{l,m} B_{lm}^r(r,t) \vec{Y}_{lm}^r(\theta,\phi)+B_{lm}^e(r,t) \vec{Y}^e_{lm}(\theta,\phi)+B_{lm}^m(r,t) \vec{Y}^m_{lm}(\theta,\phi)\,,\label{19}
\eea
with
\be
\vec{Y}_{lm}^r=\hat{r} Y_{lm}\,,\hspace{1cm}\vec{Y}_{lm}^e=(l(l+1))^{-1/2}\,  r\, \vec{\nabla} Y_{lm}\,, \hspace{1cm} \vec{Y}^m_{lm}=\hat{r}\times \vec{Y}_{lm}^e\,,\label{20}
\ee
and where $Y_{lm}$ are the ordinary spherical harmonics. 
The vector spherical harmonics form a complete orthonormal basis of vector fields on the sphere for a fixed radius. There are three types of vector harmonics: $Y^{s}_{lm}$, with $s=r,e,m$, the radial, ``electric'', and ``magnetic'' components, and there are $2l+1$ values of $m$ for each $l\ge 1$. For $l=0$ there is only the radial component.  For simplicity in what follows we will use real vector harmonics such that the coefficients in the expansion are Hermitian operators.

The constraint equations (\ref{cont}) tell the components proportional to the ``electric'' vector harmonics $\vec{Y}_{lm}^e$ (for $l\ge 1$) are dependent variables
\bea
E_{lm}^e=(l(l+1))^{-1/2} \left(2 E_{lm}^r+ r \frac{d E_{lm}^r}{dr}\right)\,,\label{tyty}\\
B_{lm}^e=(l(l+1))^{-1/2} \left(2 B_{lm}^r+ r \frac{d B_{lm}^r}{dr}\right)\,.
\eea
Therefore, the algebra is generated by the fields $E^r_{lm},E^m_{lm},B^r_{lm},B^m_{lm}$. These fields decouple for each $l,m$, $l \ge 1$, and the only components for $l=0$ are $E^r,B^r$ which identically vanish in this charge-less case. Writing the scaled variables 
\bea
\phi^1_{lm}= \frac{r^2}{\sqrt{l(l+1)}} E^r_{lm} \,,\hspace{2cm} \pi^1_{lm}=r B^m_{lm}\,,\label{310}\\
\phi^2_{lm}=\frac{r^2}{\sqrt{l(l+1)}} B^r_{lm} \,,\hspace{2cm} \pi^2_{lm}= r E^m_{lm}\,,   \label{311}
\eea
it turns out we have two independent modes given by canonical variables $(\phi^1_{lm},\pi^1_{lm})$ and $(\phi^2_{lm},\pi^2_{lm})$. From the commutation relation for the electromagnetic field (\ref{comu}), 
 it follows the two modes  have equal time canonical commutation relations as $d=2$ fields in the $t,r$ coordinates 
\bea
[\phi^i_{lm}(r,t),\pi^{i'}_{l'm'}(r',t)]= i \,\delta_{i i'} \,\delta_{l l'}\, \delta_{m m'}\,\delta(r-r') \,.
\eea  
The correlators of these Gaussian variables correspond to the fundamental state of the Hamiltonian 
\be
H=\sum_{i=1}^2\,\sum_{l\ge 1,m}\, \int_0^\infty dr\, \frac{1}{2} \left((\pi^i_{lm})^2+(\partial_r \phi^i_{lm})^2+\frac{l(l+1)}{r^2}(\phi^i_{lm})^2\right)\,,
\ee
which follows by expanding the the electromagnetic Hamiltonian (\ref{hami}).

 An expansion of a free massless scalar $\tilde{\phi}$ in spherical coordinates gives exactly the same decomposition in radial modes with the same algebra and Hamiltonian \cite{Srednicki:1993im,Casini:2015dsg}, and hence the the same correlators.  The difference is that each mode of the scalar is duplicated in the pair of variables $(\phi^1_{lm},\pi^1_{lm})$, $(\phi^2_{lm},\pi^2_{lm})$ for the Maxwell field, and that for the Maxwell field the mode $l=0$ is missing. These features are related to the helicity $1$ of the Maxwell field. Concretely, the identification is 
\bea
 \phi^1_{lm}(r,t)\leftrightarrow \phi^2_{lm}(r,t)\leftrightarrow \tilde{\phi}_{lm}(r,t)=r \int d\Omega\, \tilde{\phi}(x) \, Y_{lm}(\theta,\varphi)\,,\hspace{1cm} l\ge 1\,,\label{314}\\
 \pi^1_{lm}(r,t)\leftrightarrow \pi^2_{lm}(r,t)\leftrightarrow \tilde{\pi}_{lm}(r,t)=r \int d\Omega\, \tilde{\pi}(x) \, Y_{lm}(\theta,\varphi)\,,\hspace{1cm} l\ge 1\,.
\eea
This identification is a unitary transformation mapping operators and states.  It is  non local in space, but crucially, it is local in the radial direction, identifying algebras determined by the same arbitrary radial regions in the two theories.  
 
  Therefore, we have that the mutual information is given by twice the one of the massless scalar in $d=4$ minus twice the mutual information of the $l=0$ mode, which is  a $d=2$ dimensional scalar field with Hamiltonian 
 \be
 H= \int_0^\infty dr\, \frac{1}{2} \left(\pi^2+(\partial_r \phi)^2\right)\,,  \label{hasi} 
\ee
on the half line $r>0$, with $\phi(0)=0$ \cite{Casini:2015dsg}. This gives
\be
S_\epsilon=1/2\, I_\epsilon= k \,\frac{4 \pi R^2}{\epsilon^2}-\frac{16}{45} \log(R/\epsilon)+\frac{1}{2}\log(\log(R/\epsilon))+\textrm{const.}\,.
\ee
The coefficient $-16/45=2\times (-1/90)-2 \times (1/6)$, where $-1/90$ is the logarithmic coefficient of the scalar field, and $1/6$ the logarithmic coefficient for the $l=0$ mode (\ref{hasi}). The coefficient $k$ is universal and corresponds to the one on the mutual information between parallel planes for a scalar \cite{Casini:2015dsg}. The subleading $\log(\log(R/\epsilon))$ term comes from the mutual information of the $l=0$ mode.

\subsection{The effect of interactions}

To see how the mutual information changes with $\epsilon$ in presence of charges, as we have discussed in section \ref{physical}, we have to evaluate the change in the logarithmic term of the entropy of a thin shell when the mass $m_e$ gets smaller than $\epsilon^{-1}$. This entropy requires the introduction of a cutoff, and issues may arise, such as the precise definition of the algebra associated with the region. In a lattice, the chosen algebra might contain a center formed by operators in the boundary \cite{Casini:2013rba}. This issue is however irrelevant for the calculation we are performing because we are looking for a change in the entropy with $\epsilon$ and the possible operators localized in the boundary have large correlations with themselves in the continuum limit, such that their contribution to the entropy, whatever it is, is independent of the size of $\epsilon$.  See the discussion in section \ref{comparison}.  

Then, we expect the important physical effect of the interactions to be the change in expectation values of the smeared electric flux normal to the shell, and in turn a change in the logarithmic coefficient. But these variables form part of a larger algebra of operators in the shell, and we have to understand the variation of the quantum entropy of this algebra. 

To lowest order in QED, the effective Lagrangian is non-local but still quadratic,
\be
{\cal L}=-\frac{1}{4}\, F_{\mu\nu}\left(1+\pi(-\partial^2) \right) F^{\mu\nu}\,,
\ee
where $\pi(q^2)$ is the renormalized vacuum polarization amplitude. Therefore, we can still think in terms of Gaussian variables. 
This correction changes the equal time electric and magnetic correlators in coordinate space as 
\bea
\left\langle B_{j}\left(x\right)B_{k}\left(0\right)\right\rangle  & = &\left(\partial_{j}\partial_{k}-\delta_{jk}\nabla^{2}\right) \int_{0}^{+\infty}dm^{2}\rho\left(m^{2}\right)C^{0}\left(x,m\right)\,,\label{43}\\
\left\langle E_{j}\left(x\right)E_{k}\left(0\right)\right\rangle  & = & \left(\partial_{j}\partial_{k}-\delta_{jk}\nabla^{2}\right) \int_{0}^{+\infty}dm^{2}\rho\left(m^{2}\right)C^{0}\left(x,m\right)\nonumber \\
&& \hspace{3.7cm}+\delta_{jk}\,\int_{0}^{+\infty}dm^{2}\rho\left(m^{2}\right)m^{2}\,C^{0}\left(x,m\right)\,.\label{44}
\eea
where $\rho(m^2)$ is the spectral density (\ref{spec}), and $C^0(x,m)$ is the scalar correlator of mass $m$,
\be
C^{0}\left(x,m\right) =  \int_{\mathbb{R}^{4}}\frac{d^{4}p}{\left(2\pi\right)^{3}}\,\Theta\left(p^{0}\right)\delta\left(p^{2}-m^{2}\right)\mathrm{e}^{-ip\cdot x}=\frac{m}{4\pi^2 x} K_1(m x)\,.
\ee
The equal time commutators are kept the same. 
 
We see the electric correlator is not divergence-free any more, due to the presence of the charge density operator, and the electric-magnetic duality is broken in the absence of monopoles. These effects are due to the last term of (\ref{44}), that we naturally expect to be responsible for the non-trivial effects. This term vanishes in the decoupling limit $\alpha\rightarrow 0$. 

The constraint equation of the electric field (\ref{tyty}) is changed by the addition of the charge density operator. However, the electric component $E^e_{lm}$ is still a dependent variable, now given in terms of the radial component and the charge density. Then, in evaluating the entropy of the electromagnetic field we can restrict  our attention to the generating fields of the algebra which are the same radial and magnetic modes (\ref{310}), (\ref{311}).\footnote{In the same way, time derivatives of the fields are dependent variables through the equations of motion.} In particular, the mode $l=0$ of the radial components is given in terms of the total charge as a function of the radius. This can be thought of as a variable belonging to the charged operator algebra. Hence, for the algebra of the Maxwell field, we can still ignore the $l=0$ mode, though there is an important effect of this mode on the charged algebra that will be discussed later on in the calculation. 

The correlators of these radial variables can be readily evaluated from (\ref{spec}), (\ref{18}), (\ref{19}), (\ref{20}), (\ref{43}) and (\ref{44}). As expected, we do not get relevant changes concerning the free correlators except for the correlator $\langle E^r_{lm}(r) E^r_{lm}(r') \rangle$ of the radial electric variable, due to the last term in (\ref{44}). The perturbations for the other correlators are computed in the appendix \ref{AA}, where we also discuss why these corrections are irrelevant for the present problem.  In particular, the second mode $\phi^2_{lm}, \pi^2_{lm}$ or equivalently $B^r_{lm},E^m_{lm}$, corresponding to the radial magnetic variable does not contribute to the change in the logarithmic term. However, we expect this mode will produce a contribution in the presence of magnetic monopoles.

Therefore, we will focus on the first mode $(\phi^1,\pi^1 )$, corresponding to the radial electric field $E^r$ and the magnetic component of the magnetic field $B^m$ (\ref{310}). Let us first look at the free correlators. The scalar correlator is
\bea
&&\langle \phi^1_{lm}(r) \phi^1_{lm}(r')\rangle_0 \equiv  \frac{r^2 r'^2}{l(l+1)}\langle E^r_{lm}(r) E^r_{lm}(r') \rangle_0 \nonumber \\
&&\hspace{0cm}=\frac{r r'}{(2\pi)^2} \int d\Omega\, d\Omega'\, \frac{Y_{lm}(\Omega)Y_{lm}(\Omega')}{r^2+r'^2- 2 r \,r'\, \hat{\Omega}\cdot \hat{\Omega}'} 
= \frac{1}{4 \pi} \int d\theta\,\sin(\theta)\, \frac{P_l(\cos(\theta))}{z-\cos(\theta)}\label{322}\\
&&\hspace{0cm}= \frac{\Gamma[l+1]}{2^{1+2}\sqrt{\pi}\,\Gamma[l+3/2]}\, \frac{1}{z^{l+1}}   \,\,_2F_1\left(\frac{l+1}{2},\frac{l+2}{2},l+\frac{3}{2},\frac{1}{z^2}\right)\,,\nonumber
\eea
where 
\be
z=\frac{r^2+r'^2}{2 r r'}>1\,.
\ee
The step in the second line follows from the fact that the integral is independent of $m$ and that the spherical harmonics are eigenvectors of any rotational invariant kernel.  
Analogously, the momentum correlator reads
\bea
\langle \pi^1_{lm}(r)\pi^1_{lm}(r')\rangle_0 &=&-\frac{2 r r'}{(2\pi)^2} \int d\Omega\, d\Omega'\, \frac{Y_{lm}(\Omega)Y_{lm}(\Omega')}{(r^2+r'^2- 2 r \,r'\, \hat{\Omega}\cdot \hat{\Omega}')^2}\label{324}
\\
&=& -\frac{1}{4 \pi r r'} \int d\theta\,\sin(\theta)\, \frac{P_l(\cos(\theta))}{(z-\cos(\theta))^2}= -\frac{1}{r r'}\partial_z \langle\phi^1_{lm}(r)\phi^1_{lm}(r')\rangle_0\,.  \nonumber
\eea

In the thin shell $r\in (R-\epsilon/2,R+\epsilon/2)$ we have $|r-r'|/R\ll 1$ and the correlators behave as the one for a $d=2$ scalar,  
\bea
\langle \phi^1_{lm}(r) \phi^1_{lm}(r')\rangle_0 &\sim & -\frac{1}{2\pi}\log(|r-r'|/R)\,,\label{freee}\\
\langle \pi^1_{lm}(r)\pi^1_{lm}(r')\rangle_0 & \sim & -\frac{1}{2 \pi \, |r-r'|^2}\,.\label{freep}
\eea	
These limits can be more simply understood by noting that the integrals (\ref{322}) and (\ref{324}) are dominated for small $|r-r'|/R$, $z\sim 1$, by $\theta\sim0$, where we can  replace $P_l(\cos(\theta))\sim P_l(1)=1$. This behavior, independent of $l$,  persists while $l\ll R/|r-r'|\sim R/\epsilon$. For larger angular momentum, the oscillatory dependence of the Legendre function changes the result. The full tower of $l$ in the interval gives the scalar entropy in the shell, but we will focus on the  modes of low $l$ which are the responsible for the change in the logarithmic term. 

Except for unimportant corrections discussed in appendix \ref{AA}, the only relevant one to these correlators is for the radial electric field and is due to the last term in (\ref{44}). This term gives  
\bea
&&\Delta \langle \phi^1_{lm}(r) \phi^1_{lm}(r')\rangle \equiv  \frac{r^2 r'^2}{l(l+1)}\Delta \langle E^r_{lm}(r) E^r_{lm}(r') \rangle\nonumber
\\
&&= 
\frac{r^2 r'^2}{l (l+1)} \int d\Omega\, d\Omega'\, Y_{lm}(\Omega)Y_{lm}(\Omega') (\hat{\Omega}\cdot \hat{\Omega}') \Delta C\left(\sqrt{r^2+r'^2- 2 r r' \hat{\Omega}\cdot \hat{\Omega}'}\right)\,,\label{326}
\eea
where 
\be
\Delta C(x)= \int_{0}^{+\infty}dm^{2}\rho\left(m^{2}\right)\, m^2 \, C^{0}\left(x,m\right)\,.\label{327}
\ee
This new term contains the effect on the normal fluxes and will be the responsible of the change in the logarithmic term. 
The function in (\ref{327}) is exponentially small for  $m_e x\gg 1$, and for $m_e x\ll 1$ we have 
\be
\Delta C(x)\sim  \frac{\int_{4 m_e^2}^{1/x^2} dm^2\, m^2\,\rho(m^2)}{\,x^2} 
 \,.
\ee
Note the UV behavior depends on the spectral function. For QED at the lowest order,  it gives 
\be
 \Delta C(x)\sim  \frac{\alpha}{ 3 \pi^3  \, x^4}\,.\label{eq}
\ee
 The precise behavior will not be relevant as far as it dominates over the free contribution for small $x$. This implies a spectral density falling slower than $\rho(q^2)\sim q^{-4}$ for large $q^2$.  This coincides with the condition that the fluxes get an area term diverging for small $\epsilon$, and the unitarity bound for the current correlators in a scaling limit, as discussed in section \ref{physical}.
 
For (\ref{eq}) eq. (\ref{326}) gives
\be
\Delta \langle \phi^1_{lm}(r)\phi^1_{lm}(r')\rangle\sim \frac{ \alpha}{3 \pi^2} \,(l (l+1))^{-1} \frac{R^2}{ |r-r'|^2}\,.\label{correc}
\ee
 This again is valid for $l\ll R/\epsilon$, independently of the mass, as far as we are in the regime $m \epsilon\ll 1$.   
 Notice that due to the tensor structure of the second term in (\ref{44}), as opposed to the first term in the same equation, the $l(l+1)$ dependence coming from the normalization of the radial electric field does not disappear for this correction. This factor encapsulates the main effect affecting the statistics of the modes $l\lesssim R/\epsilon$, and displays the phenomenon of enlarged self correlations for the smeared electric fluxes now written in terms of the angular modes. There will be changes for large angular momentum $l\ge R/\epsilon$ too, but these are local, and would not modify the mutual information. In fact, the contribution to mutual information falls exponentially fast for $l> R/\epsilon$ because $l/R$ plays the role of a mass in a picture of dimensional reduction with respect to the  directions parallel to the surface, and correlations between the two regions on both sides of the shell are exponentially suppressed for $\epsilon$ larger than the mass.

Now, let us recall the formula for the entropy of Gaussian variables with correlation kernels $X$ and $P$ for the field and the momentum variables,  
\be
S=\textrm{tr}\left((\sqrt{ X P} +1/2) \log(\sqrt{X P }+1/2)-(\sqrt{ X P} -1/2) \log(\sqrt{X P }-1/2)\right)\,.
\ee
For the regime of low angular momentum $l$, the state in the interval is very entropic because the product of correlation functions is large. For example, $\textrm{tr}(X P)\sim \alpha R^2/\epsilon^2 /l^2\gg 1$. Therefore we can safely discard the $1/2$ inside the logarithms in the above formula to approximate for each mode
\be
\Delta S_l= \textrm{tr} \log(\sqrt{X_l P_l })=-\frac{1}{2}\log(l(l+1)R^{-2})+\textrm{const}\,,\label{firs}
\ee
where the constant is the entropy given by the correlators (\ref{freep}) and (\ref{correc}) without the $l$ and $R$ dependent factors in this later formula, and subtracted from the one of the free scalar. This later is an $l$ independent entropy of a $d=2$ model in an interval. The important point is that it does not have a dependence on $l$ and its contribution summed over the spherical modes is proportional to the trace of an identity operator on the sphere, which will add a contribution to the area term. 

The entropy produced by the first term in (\ref{firs}) can then be written as 
\be
\Delta S=-\frac{1}{2}\textrm{tr} \log (-\nabla_\Omega^2)\,,\label{entro}
\ee
where the operator inside the logarithm is the Laplacian on the sphere of radius $R$. The mode $l=0$ is absent in the definition of the Laplacian. The size of the regularization we have to impose on expression (\ref{entro}) is precise, we have a distance cutoff $\epsilon$ in the sphere, corresponding to the limit on the angular momentum, $R/l > \epsilon$. 

This calculation can be done by standard methods, for example using the heat kernel. The heat kernel is defined as
\be
K(\tau)= \textrm{tr} \,e^{-\tau (-\nabla_\Omega^2)}=\sum_{l \ge 0} (2 l+1) e^{- \frac{\tau}{R^2} \, l (l+1)}- 1 \,,\label{335}
\ee
where we have subtracted the mode $l=0$. For small $\tau$, using Euler MacLaurin formula (see for example \cite{Huang:2014pfa}), we have
\be
K(\tau)\sim \frac{R^2}{\tau}+\frac{1}{3 }-1+{\cal O}(\tau)\,.
\ee
The trace in (\ref{entro}) follows from the formula\footnote{This formula tells us the $\log \epsilon$ coefficient, which gives us also the $\log R$ coefficient because of scale invariance. If we have kept the zero mode we would need an infrared regulator, for example a small mass $\mu$. Then there would be an additional contribution $-\log(\mu \epsilon)$ to (\ref{oop}). We would have obtained $-1/3$ for the coefficient of $\log(\epsilon)$ compensated by different coefficients for $\log R$ and $\log \mu$.  }
\be
-\frac{1}{2}\, \textrm{tr}\log( -\nabla_\Omega^2)=\frac{1}{2} \int_{\epsilon^2}^\infty \frac{d\tau}{\tau}\, K(\tau) =
\textrm{area term}+\frac{1}{3} \log(R/\epsilon)- \log(R/\epsilon)+\textrm{cons}\,.\label{oop}
\ee
We have kept separated the contribution of the (absent) mode $l=0$ because it will soon be canceled by a different term.  

Eq. (\ref{oop}) gives the change in the logarithmic term of the entropy of the shell. It goes with a negative sign in the mutual information, that changes as
\be
\Delta I_{\textrm{Maxwell}}=I_{\textrm{interacting}}-I_{\textrm{free}}\sim \cdots -\frac{1}{3} \log(R/\epsilon)+ \log(R/\epsilon)+\cdots \label{cot}
\ee

There is also a contribution to the mutual information of the charged fields. As they are very massive again the naive expectation is that there is no $\log R$ term coming from this sector. However, there is a constraint in the algebra of the charged fields in the sphere or its complement. Only neutral operators appear in these algebras because they are the only operators that are local when interacting with the Maxwell field. Then, the algebra of the charged fields is, in fact, a $U(1)$ orbifold. See \cite{Pretko:2018yxl,Zuo:2016knh} for previous discussions where this contribution of charged particles to the Maxwell field entropy was recognized. As discussed in \cite{Casini:2019kex}, there is a universal logarithmic correction to the mutual information for these orbifolds that shows up, even for very massive fields, once $\epsilon m_e\ll 1$ . This is given by
\be
 I_{\textrm{orbifold}}-I_{\textrm{full}}=- \frac{d-2}{2}\log(R/\epsilon)+\cdots   \label{estotra}
\ee
Here $I_{\textrm{full}}$ is the mutual information for the algebra of the full charged massive fields, which does not contain any logarithmic term. We review this result from the perspective of the replica calculation of the entropy in appendix \ref{orbi}.   
For $d=4$, this exactly cancels the contribution of eliminating the $l=0$ mode in (\ref{cot}). This is no coincidence. The contribution in (\ref{estotra}) comes from the entropy of total charge (Gaussian) fluctuations in the sphere (which are compensated in the complement). This entropy is subtracted in the orbifold \cite{Casini:2019kex}. This entropy is equal through Gauss law to the one associated with the total electric flux fluctuations in the shell, corresponding to the $l=0$ mode. This contribution could then be used to complete the Laplacian on the sphere with the mode $l=0$ with a specific infrared cutoff $\sim R$. If we have kept this contribution in the above calculation of the shell entropy of the electromagnetic field it would also be subtracted in the mutual information, as it is subtracted in the orbifold mutual information. 
Hence, alternatively, we could have considered the radial $l=0$ flux as part of the Maxwell field algebra and not correct for the zero mode in (\ref{335}), while at the same time disregard the fluctuations of the total charge operator in the charged field algebra, which is the one that makes a difference for the orbifold.

In conclusion, we have a $-1/3 \log(R/\epsilon)$ correction for the mutual information, that goes into the regularized entropy with an additional factor of $1/2$. Therefore, for the Maxwell field interacting with electric charges,
\be
S_{\textrm{reg}}^{\textrm{int}}=\cdots -\left(\frac{16}{45}+\frac{1}{6}\right) \log(R/\epsilon)+\cdots
\ee
which still does not match the anomaly. 

Interestingly, to get the anomaly one has to consider the effect of monopoles. They will affect the dual modes $E^m,B^r$, containing the radial magnetic fluxes. The correction is thus duplicated  
\be
S_{\textrm{reg}}^{\textrm{int}}=\cdots -\left(\frac{16}{45}+\frac{1}{6}+\frac{1}{6}\right) \log(R/\epsilon)+\cdots
=\cdots -\frac{31}{45} \log(R/\epsilon)+\cdots
\ee
having the right anomaly coefficient. 

The necessity to invoke monopoles might be surprising. However, it is completely natural from the fact that the problem to solve was for the free Maxwell field in the IR and this is a duality invariant problem. It is also necessary when considering RG flows. One starts with the mutual information for the Maxwell field in the IR with the hope that decreasing $\epsilon$ one would get the right anomaly by adding the effect of charges. If the electric charges at some scale would solve the problem and provide the right anomaly, we would be into another problem. This is because in the deep UV the theory might contain also monopoles which would then spoil the matching with the anomaly when crossing that new scale. The existence of monopoles seems necessary to have a complete theory with quantized electric charges.  

\subsection{Comments on the literature}
\label{comparison}

The subject of EE in gauge theories has attracted much attention in the literature. One issue that was much discussed is how to split the Hilbert space as a tensor product for complementary regions. In a lattice gauge theory, gauge dependent variables are assigned to links. A tensor product decomposition across a boundary can be implementated by the construction of an extended lattice with new special vertices, not associated to gauge transformations,  at the points where the boundary cuts a link \cite{Buividovich:2008gq,Donnelly:2008vx,Donnelly:2011hn}. Another implementation, an "extended Hilbert space" approach, defines an enlarged Hilbert space for non gauge invariant fields, while keeping the state gauge invariant \cite{Ghosh:2015iwa,Soni:2015yga}.  However, the EE in lattice gauge theory has a natural definition as the entropy of a state in an algebra of local gauge invariant operators \cite{Casini:2013rba}. This definition is in fact the same as for any other model; entropy in quantum mechanics is the entropy of a state in a particular algebra, and the entropy of a region is the one of an algebra of operators attached to it. Issues may arise in a lattice concerning the precise algebra assigned to a region. The entropy for both, the extended lattice and the extended Hilbert space approaches, corresponds to a particular choice of local algebra called the electric center choice in \cite{Casini:2013rba}. This consists on all gauge invariant operators in the region plus the electric field normal to the boundary. This electric field commutes with the rest of the algebra and forms a center for it. The entropy contains a classical Shannon piece due the presence of this center. There are infinitely many other possible choices of local algebras that differ by details on the boundary, in particular there are many choices without center, and hence defining a tensor product decomposition. The entropies of all these choices differ in the same way that entropies for different regularizations differ to each other. In the continuum limit, the quantities that are well defined and finite for QFT such as the relative entropy and mutual informations, are independent of these particular choices \cite{Casini:2013rba}. See \cite{Donnelly:2015hxa,Donnelly:2014fua,Soni:2016ogt,VanAcoleyen:2015ccp,Donnelly:2014gva,Aoki:2015bsa,Radicevic:2015sza,Hung:2015fla,Moitra:2018lxn,Radicevic:2016tlt,Donnelly:2016mlc} for further developments.  

In  \cite{Donnelly:2015hxa,Donnelly:2014fua,Soni:2016ogt} it was argued that for a free Maxwell field it is precisely the electric center (or ``edge modes") classical term that produces a contribution to the logarithmic term that restores the anomaly coefficient.\footnote{Negative contributions to the area term have also been discussed, see for example \cite{Donnelly:2012st,Kabat:1995eq,Solodukhin:2012jh}.}
 This contribution is given by the classical entropy of the electric field normal to the sphere on the boundary. See also \cite{Huang:2014pfa} where this same contribution is attributed to gauge modes at the boundary. The solution discussed in this paper also depends on the statistics of the normal electric (and magnetic) fluxes near the boundary, and both calculations end up with the partition function of a scalar on the surface of the sphere, eq. (\ref{entro}).   
In a certain sense, our paper gives a justification for the technical result of these previous calculations. However, we want to highlight several important differences. 

The problems posed by the idea of the contribution of a center term in QFT have not been much appreciated. 
 In general local algebras in the continuum theory do not contain a center. To commute with the rest of the algebra an operator has to be localized in the boundary, and it is not possible to localize an operator in a surface of $d-2$ dimensions. Such operator would be too singular to be an operator in Hilbert space, in the same way field operators at a point are not Hilbert space operators but operator valued distributions. In terms of a lattice model, this means these operators will tend to have very large self-correlations and decouple from the rest in the continuum limit. That is why they do not affect the mutual information. 
 In this sense, the results of \cite{Donnelly:2015hxa,Donnelly:2014fua,Soni:2016ogt} highlight that the ambiguities in the entropy also reach to the logarithmic term for some regularizations.  This emphasizes the importance   to use a quantity that remains physical in the continuum to settle this issue. This is the case of the mutual information. For the bare entropy, the electric center is a particular choice, and other choices will produce different results. As we have shown, exactly the same electric center choice for the Maxwell field can be mapped to a center choice for a scalar theory giving ambiguities also in this case. The correlators for the radial electric field $E^r_{lm}$ coincide with the ones of the scalar modes $R^2 \sqrt{l(l+1)}\phi_{lm}$. It is interesting to notice that the effect on the logarithmic term will appear in the scalar representation due to the factor depending on $l$, and this is only relevant because of the classical entropy of continuum variables is not well defined, and is not invariant under changes of normalization. This emphasizes the ill defined nature of these contributions. If we include in the algebra the conjugate momentum along with the radial electric field, the normalization is automatically irrelevant, and the result for the free Maxwell field is equivalent to the scalar one (minus the $l=0$ mode), with no additional logarithmic contribution. There is also an important point in the calculation of the contribution of the electric center for the free field. The total flux for the free field is zero, and then the mode $l=0$ should be absent in evaluating the spectral quantity (\ref{entro}). This gives a correction to the entropy $2/3\log(R/\epsilon)$ instead of $-1/3\log(R/\epsilon)$, and the result does not match the anomaly.         
 
Our results for the mutual information, which are free from ambiguities, also rely on a surface effect, but  charges are crucially necessary for this effect to take place, and the mass of these charges sets the scale of the surface width. The importance of taking into account charges when computing the entropy of a Maxwell field was also emphasized in \cite{Pretko:2018yxl,VanAcoleyen:2015ccp}.  The result for a free Maxwell field is not the anomaly coefficient. We compute quantum entropies, and the effect given by eq. (\ref{entro}) is not a classical entropy but the result of an approximation in which the state is in the classical regime of large entropy because of the   large electric (and magnetic) flux fluctuations.   
 
Another important conceptual remark that underlies the present work is that there is nothing intrinsically different for models described by gauge fields in QFT that requieres a special treatment for the EE. As we have argued, the particular problem for the Maxwell field is due to the fact that in the IR it possess certain constraints that are relaxed by the UV physics. A somewhat simpler realization of an unprotected RG charge in the IR occurs for orbifolds \cite{Casini:2019kex}. As we have seen, this is also relevant to get the right anomaly coefficient for the interacting Maxwell field. The same phenomenon also happens for topological models. There are some works in the literature that link the supposed existence of a center entropy for gauge fields with the origin of the area term of holographic EE in the bulk \cite{Lin:2018bud,Lin:2017uzr,Donnelly:2016auv}. In the holographic case, what seems again to be going on is rather a macroscopic physical phenomenon which connects the UV with the IR as in models with supeselection sectors \cite{Casini:2019kex}. 
 
As we understand, the numerical result of the calculations in \cite{Donnelly:2015hxa,Donnelly:2014fua,Soni:2016ogt} would not match the anomaly if the absence of the mode $l=0$ for the free field would have been properly taken into account.  
Disregarding this point, we also find that the correction giving the anomaly is related to a partition function of a Laplacian on the $S^2$ sphere. See also \cite{Huang:2014pfa}.  However, our result comes from a very different computation. The differences at the technical level can be summarized by the equation
\be
 2\times \frac{1}{2}\left[(-)(-)(-1/3 +1) -1\right]=-1/3\,. \label{cuenta}
\ee    
The electric center for the free Maxwell field is supposed to give an entropy which is added with positive sign to the entropy of the sphere and gives a $-1/3$ logarithmic coefficient. We claim the $l=0$ mode is not present in the description of the independent variables of the free Maxwell field, what adds $1$ to the coefficient, and this should be the correct result of an electric center correction to the free field. In our setup, the effect appears for the full quantum algebra of the interacting field in the shell rather than the classical algebra of the free electric field, and is a destruction of correlations with respect to the free Maxwell field, what gives minus sign, getting $(-)(-1/3 +1)$. However, we have found the effect in the shell entropy, which appears with a minus sign in the mutual information, hence the second minus sign in (\ref{cuenta}).  
 The additional term $-1$ inside the square brackets in (\ref{cuenta}) comes from the logarithmic contribution of the charged fields. The algebra of charged fields is restricted to contain only neutral operators in the sphere. These are the only operators that can be localized due to the coupling with the Maxwell field, disregarding the size of the coupling. This constraint produces the logarithmic term for the charged field sector.   
There is also a global factor $1/2$ that comes from the regularized entropy in terms of the mutual information. This is overcome by the effect of magnetic monopoles, which is identical to the one of electric charges, and gives a factor $2$.  Therefore, the solution is explicitly electromagnetic duality invariant, and the use of mutual information is very important to clarify that. 

\section{Why should the coefficient for the interacting field coincide with the anomaly?}

\label{why}

In the previous section we started from the knowledge of the logarithmic coefficient for the free Maxwell field,  and followed the changes in the mutual information as the parameter $\epsilon$ crosses the scale of electric and magnetic charge fluctuations. In this way, we arrived at a coefficient $-31/45$, coinciding with the anomaly, for a Maxwell field interacting with heavy electric and magnetic charges. In this section, we follow the inverse direction: we will first argue that the logarithmic coefficient should be the anomaly for a complete theory, and from there we will attempt to arrive at the result for the free Maxwell field. 

Let us first review the derivation of the coefficient of the logarithmic term in the entropy for a CFT by mapping the sphere to de Sitter space. 
This is straightforward \cite{Casini:2011kv}. We conformally map the causal domain of dependence of the sphere of radius $R$ to the static patch in de Sitter space of curvature scale $R$. The vacuum state is mapped into the de Sitter symmetric vacuum state which has a specific temperature $T=(2 \pi R)^{-1}$ associated with the de Sitter Hamiltonian.  The EE of the sphere in Minkowski space is mapped to the thermodynamic entropy in de Sitter space. This is given by
\be
S=\beta E + \log (Z)\,.  \label{fore1}
\ee
The energy density is finite, and,  as the volume of the static patch is finite, the expectation value of the energy $E$  does not contribute to the divergent logarithmic term. The logarithmic term is then just given by the logarithmic term in $\log Z$, that, for de Sitter space at this particular temperature, is the free energy in the Euclidean sphere $S^d$. This gives the standard result
\be
S_{\cal F}=\cdots + (-1)^{\frac{d}{2}-1}\, 4\, A \, \log (R/\epsilon)\,. 
\ee

This derivation involves the bare entropy. It is supposed that with a local and geometric cutoff this result cannot be modified. However, as we have explained above, this can be challenged if we can modify the content of the regularized algebra with operators in the boundary such that these operators have sufficiently non-local correlations along the surface. Any change in regularization along the surface introduces boundary objects in the partition function on de Sitter space,  breaking the de Sitter invariance of the calculation. The question is when these changes can modify the RG charge.\footnote{Note that a regularized version of the electric center for the Maxwell field would contain exponentials $e^{i \lambda n \Phi_E}$ of smeared electric fluxes $\Phi_E$ on different patches with coefficients proportional to integers $n$, such that these operators close an algebra. The entropy of this classical discrete subalgebra is well defined. The scalar version of this algebra is very non-local along the surface.}  

To clarify the situation we use the mutual information for small $\epsilon$. We can think in two cases where the shell entropy contains non-local contributions. The first is a model with global SS. This corresponds to a subalgebra of a complete theory with a global symmetry group $G$. The subalgebra contains all operators that are invariant under the symmetry (an orbifold). In that case, the shell algebra contains the twist operators, that implement the symmetry only inside the sphere and not outside of it. The twist operators are non-local since they cannot be generated locally by field operators in the shell.  The second case is when there are gauge SS. In this case, there are charge measuring operators, fluxes of electric and magnetic fields, or more generally Wilson loops and t'Hooft loops. These are locally generated in the shell but must have perimeter law fluctuations because of the absence of charges. 

In a complete model, the twist operators cannot belong to the algebra of the shell since they do not commute with the charged operators in the sphere. For the case of a complete gauge theory, the sharp electric and magnetic fluxes inside the shell have area law expectation values. Then, we expect that for complete models no local changes in the regularization could challenge the result for the logarithmic term in the smooth sphere partition function, and this should coincide with the anomaly for the Maxwell field  \cite{Cappelli:2000fe,DeNardo:1996kp,Fursaev:1996uz}.  

For non-complete models, the proof using the mapping to de Sitter space should be essentially correct, but the result can change depending on the detail of the objects we insert at the boundary or the possible non-local correlations of these objects. This implies there are ambiguities in the entropy which go beyond the usual local UV ambiguities and have a more physical origin. The mutual information resolves these ambiguities. 

To understand how these non-local contributions appear for incomplete models in the mutual information let us think in the replica twist operators. The Renyi entropy of the shell for integer $n$ is given by the logarithm of an expectation value,
\be
S_n=(1-n)^{-1}\log \langle \tau_n(0) \tau_n^\dagger(\epsilon) \rangle\,, \label{qq}
\ee
where the theory is now the $n$ replicated model, and the Renyi twist operators are seated at the two boundaries of the shell, implementing the cyclic gluing of copies \cite{Calabrese:2004eu}. An OPE of the product of twists in (\ref{qq}) should contain a combination of all possible operators in the shell with the quantum numbers of the vacuum. In the short $\epsilon$ limit, the OPE should be dominated by products of operators acting on each copy of the replica manifold.\footnote{See \cite{Headrick:2010zt,Cardy.esferaslejanas,Bousso:2014uxa,Casini:2017roe} for other uses of OPE of replica twist operators.} In the limit $n\rightarrow 1$ this leaves us with expectation values of operators in the single copy theory. But these operators must belong to the shell algebra and, generally, they should not pose a problem for the RG charges. 

However, in an orbifold, the OPE contains and an additional factor of the twists operators averaged over the group. This is allowed in the shell since they commute with the uncharged operators in the ball. We show this in more detail in appendix \ref{orbi}. The result is 
\be
S_{\textrm{orbifold}}=S_{\textrm{full}}+\frac{1}{2} \log \left(|G|^{-1}\sum_{g\in G} \langle \tau_g \rangle\right)\,.  \label{dion}
\ee
The $\tau_g$ are twist operators seated in the shell, with typical smearing of size $\epsilon$, and the $1/2$ factor comes from the mutual information regularization. Taking into account the statistics of the expectation values of sharp twists, this gives, for example, the contribution (\ref{estotra}) for a $U(1)$ orbifold \cite{Casini:2019kex}. This is a zero (modular) temperature contribution to the entropy since the correction does not depend on the Renyi index, see appendix
 \ref{orbi}. However, for a massive field, it comes from correlations at a distance $\sim m^{-1}$ at both sides of the boundary.   

Notice that the correction is just an average over the possible non-local operators on the shell. Other operators may contribute but do not give a non-local contribution that changes the RG charge. In this scenario with global symmetries, we have two models, where the algebras contain or not charged operators, and this leads to two different results. Let us now think in the case of the complete theory of a Maxwell field with charges. We again expect to have an analogous contribution to the entropy given by sums over operators on the shell. The important part of the contribution that would contain the non-local correlations should be, in analogy with (\ref{dion}),
\be
S_{\textrm{log}}=\log Z(S^{4})+\frac{1}{2}\log \left(N^{-1}\sum_{\Gamma,\Gamma',q,g} \langle e^{i (g \int_\Gamma E_r+q \int_{\Gamma'} B_r}\rangle \right)\,.
\ee
Here $\Gamma,\Gamma'$ are patches on the shell, $q$, $g$ are arbitrary charges, and $N$ is a normalization factor.
 
This should not produce corrections to the logarithmic term as far as the flux operators have an area law. Once we have increase $\epsilon$ enough to have free field expectation values for the smeared loop operators the situation changes. These fluxes can then be written as Wilson and t'Hooft loops on the shell having perimeter law expectation values, 
\be
S_{\textrm{log}}=\log Z(S^{4})+\frac{1}{2}\log \left(N^{-1}\sum_{\Gamma,\Gamma',q,g} \langle W_\Gamma^q  T_{\Gamma'}^g\rangle \right)\,.
\ee
 We can write in an effective way the new contribution as a path integral on the boundary $\Sigma$
\bea
 && \frac{1}{2}\log \langle \int {\cal D}\alpha \, {\cal D}\beta\, e^{i \int_\Sigma d\sigma\, (\alpha(x) E_r(x)+  \beta(x) B_r(x))}\rangle \nonumber \\
 &&= \frac{1}{2}\log \int {\cal D}\alpha \, {\cal D}\beta\, e^{-\frac{1}{2} \int d\sigma_1\, d\sigma_2\, (\alpha(x_1)\langle E_r(x_1)E_r(x_2)\rangle  \alpha(x_2)+\beta(x_1)\langle B_r(x_1)B_r(x_2)\rangle  \beta(x_2))} \,,\label{sty}
\eea
where the regularization scale is set to $\epsilon$, and the integrals are normalized $\int {\cal D}\alpha=\int {\cal D}\beta=1$. This gives the contribution 
\be
-\frac{1}{4} \textrm{tr} \log (G_E)-\frac{1}{4} \textrm{tr} \log (G_B)\,,
\ee
where $G_E$ and $G_B$ are the radial electric and magnetic correlator kernels on the surface. The calculation of this type of contributions was done \cite{Soni:2016ogt} in the context of the electric center contribution to the entropy.\footnote{As discussed in the previous section, with respect to the calculation in \cite{Soni:2016ogt}, we have a difference in an additional factor $1/2$ because of the mutual information regularization, compensated by the addition of the magnetic fluctuations on top of the electric ones. There is also a global sign $-1$ since we are not computing the entropy of the electric fluctuations but just the partition function (\ref{sty}), and this contribution is not part of the full coefficient $-31/45$ but an additional piece that is added for large $\epsilon$. We also have to make the same comments as in the previous section about the mode $l=0$. This flux is set to zero with no fluctuations in the free Maxwell field, but is compensated by the loss of the contribution of the orbifold of the charged sector as we move to large $\epsilon$.}  The result for the universal piece is $\frac{1}{3}\log(R/\epsilon)$. Therefore, starting from the logarithmic coefficient $-31/45$ for short $\epsilon$ in the complete theory we again arrive to $-16/45$ for larger $\epsilon$,  which corresponds to the pure free Maxwell field.

\section{Final remarks}
\label{conclusion}

We have shown the mismatch of the logarithmic coefficient of a free Maxwell field is solved by the presence of electric and magnetic charges, as far as the regulating distance is set to be smaller than the typical mass scale of the charge fluctuations. 

The reason for the mismatch for the free Maxwell field is the existence of certain operators, electric and magnetic fluxes, with peculiar long-distance correlations. This leads to some degree of non-protection of the infrared RG charge. However, this is not relevant for the irreversibility theorems since the coefficient for a complete model is always the same in the limit of vanishing regulator and large radius.  The phenomenon does not have a relation with gauge symmetries, but with the existence of superselection sectors in the IR theory. A similar phenomenon exists for other models with SS sectors.      
 Models without IR superselection sectors do not display these types of alternatives. 
 
The effect of the IR SS on the entropy cannot be described as a pure UV nor a pure IR phenomenon. It is rather an effect on the IR entropy facilitated by UV physics. The main witnesses of this physics are the smeared flux operators (Wilson and t'Hooft loops) that sense both the UV and the IR by having a large size along the surface and a short one in the perpendicular direction.  

Through this paper, we have analyzed the case of an IR free Maxwell field interacting with heavy charges. The matching with the anomaly will also hold for asymptotically free gauge theories and regions of size $R$ in the UV regime, where the theory is complete in the sense that it contains charges for all representations. The full anomaly (without orbifold corrections) has to be assigned to the charged fields. In this regime, we do not have the constraint that $\epsilon$ should be smaller than a mass scale, but $\epsilon$ should be small enough to satisfy (\ref{yestaquetal}). This is achieved with $\epsilon\ll  \alpha(M)\, R /|\log (R M)|$, with $M$ the confinement scale.  

Previous discussions in the literature about this subject give the correction in the entropy as a classical entropy of a center in the algebra, and this piece is supposed not to quantify entanglement but just classical correlations.  We can wonder if our results describe the correction to the entropy as a quantum or a classical contribution. Our discussion was in terms of mutual information, to deal with well-defined quantities. This does not allow us to discern if there are classical correlations or, for example, distillable entanglement. An answer to this question in any QFT requires to look at different measures of entanglement instead of the mutual information \cite{Hollands:2017dov}. At present, this seems very hard in QFT. For a finite system in a pure state, all the natural measures of entanglement agree (for algebras without center) with the entanglement entropy. In a general QFT, we do not know if the expansions of the different entanglement measures with the separation distance agree all the way to the universal coefficient. But given that the anomaly is obtained for the complete model, we can expect that the answer to the question about the amount of the entropy that can be considered classical or quantum would not differ qualitatively from the one for simpler models such as a free scalar.   

A final important remark is that we have found an interesting and simple effective way of describing the contribution of IR superselection sectors to the entropy, that applies to both, global and local superselection charges. The formula consists of the logarithm of the average of expectation values of operators that contribute to the non-local correlations along the surface. Recently \cite{Casini:2019kex}, we have proposed that holographic theories should be thought of as theories having a large number of effective superselection sectors. The contribution of these sectors to the entropy should give the dominant bulk area term to the holographic entropy. This results in an interesting perspective that the Ryu-Takayanagi formula may correspond in the boundary QFT to an average of expectation values over a large set of surface operators of the theory.

\appendix

\section{Other corrections to correlation functions of spherical modes}
\label{AA}
In this appendix, we analyze the corrections for the correlators of the radial variables other than the radial electric mode (in the absence of monopoles). 

Let us analyze  first the second mode $\phi^2_{lm}, \pi^2_{lm}$ or equivalently $B^r_{lm},E^m_{lm}$. 
The non trivial spectral density in (\ref{43}) will affect the correlations of $B^r$ at short distance but will not introduce important qualitative differences since these corrections keep the correlators divergenless. To convince ourselves of this statement we can again look at the fields decomposed in vector spherical harmonics and compare the theory in the sphere with a scalar one.  Writing a new two point function for a scalar $\tilde{\phi}$ as 
\be
\langle \tilde{\phi}(x)\tilde{\phi}(0)\rangle =C(x)= \int_{0}^{+\infty}dm^{2}\rho\left(m^{2}\right)C^{0}\left(x,m\right)\,,\label{coco}
\ee
 we get for the correlator of the scalar spherical modes  (see \ref{314})
\bea
&& \langle \tilde{\phi}_{lm}(r)\tilde{\phi}_{lm}(r')\rangle =r r' \int d\Omega\, d\Omega'\, Y_{lm}(\theta,\varphi)Y_{lm}(\theta',\varphi')\,C(|x-x'|) \nonumber\\
&=& \frac{r r'}{Y_{l0}(0)}\int d\Omega\,\,Y_{l0}(\theta) \, C(\sqrt{r^2+r'^2-2 r r' \cos(\theta)})\nonumber \\
&=& -\frac{r r'}{l(l+1)Y_{l0}(0)}\int d\Omega\,\,(\nabla^2_\Omega \,Y_{l0}(\theta)) \, C(\sqrt{r^2+r'^2-2 r r' \cos(\theta)})
\nonumber\\
&=& -\frac{r r'}{l(l+1)Y_{l0}(0)}\int d\Omega\,\, \,Y_{l0}(\theta) \,\left(\partial_\theta^2+\cot(\theta)\, \partial_\theta\right) C(\sqrt{r^2+r'^2-2 r r' \cos(\theta)})\,\nonumber
\\&=&\frac{r^2 r'^2}{l(l+1)}\int d\Omega\, d\Omega'\, Y_{lm}(\theta,\varphi)Y_{lm}(\theta',\varphi')\, \hat{x}_i \,\hat{x}'_j\,\left(\partial_{i}\partial_{j}-\delta_{ij}\nabla^{2}\right)C(\left|x-x'\right|)\nonumber\\
&=&\frac{r^2 r'^2}{l(l+1)}\langle B^r_{lm}(r) B^r_{lm}(r') \rangle=\langle \phi^2_{lm}(r) \phi^2_{lm}(r') \rangle \,.\label{large}
\eea
The first and last steps follow from the fact that the spherical harmonics are eigenvectors of any rotational invariant kernel in the sphere and the eigenvalues do not depend on $m$. For the free case, this identification is of course the same discussed in section \ref{frito} in terms of radial Hamiltonians. An analogous calculation gives for the correlators of the magnetic components of the electric field 
\be
r r'  \langle E^m_{lm}(r) E^m_{lm}(r') \rangle=\langle \dot{\tilde{\phi}}_{lm}(r)\dot{\tilde{\phi}}_{lm}(r')\rangle=\langle \tilde{\pi}_{lm}(r)\tilde{\pi}_{lm}(r')\rangle\,.
\ee
This shows the identification (\ref{311}) of the mode $B^r_{lm},E^m_{lm}$  with a scalar mode $\phi^2, \pi^2$ for $l\ge 1$ persists. The entropy and mutual information of this mode is then equivalent to the one of a scalar interacting with heavy particles with correlator (\ref{coco}). We do not expect this to produce a change in the IR logarithmic coefficient. The possible non local changes in the entropy of the shell are determined by the low angular momentum modes $l\ll R/\epsilon$ for which the change in the correlation function is independent of $l$ and, as we have discussed in the main text, will lead to changes in the area term. In QED this correction for small $l$ is a logarithmic correction $\Delta \langle \phi^2_{lm}(r) \phi^2_{lm}(r') \rangle \sim \alpha \log^2(|r-r'|/R)$ which has to be resumed with the RG for very small $|r-r'|$.   

The correction for the magnetic component $B^m_{lm}$ which acts as a conjugate momentum of $E^r$ is again independent of $l$. A direct calculation similar to (\ref{large}) gives  
\be
\langle \pi^1_{lm}(r)\pi^1_{lm}(r')\rangle =r r'\langle B^m_{lm}(r) B^m_{lm}(r')\rangle=- \nabla^2 \int d \Omega\, d\Omega'\, Y_{lm}(\Omega)\, Y_{lm}(\Omega')\, C(|x-x'|)\,.
\ee
For small $|r-r'|$ we get an unimportant logarithmic perturbative correction to (\ref{freep})
\be
\Delta \langle \pi^1_{lm}(r)\pi^1_{lm}(r')\rangle \sim -\frac{\alpha}{6 \pi^2} \log\left|\frac{r-r'}{R}\right|\frac{R^2}{|r-r'|^2}\,.
\ee

\section{Replica trick for orbifolds}
\label{orbi}

The EE for neutral subalgebras under the action of a global symmetry group was treated in detail with an operator algebra approach in \cite{Casini:2019kex}. Here, we explicitly do the calculation of the mutual information in the coincidence limit using the replica method. 
 
Consider a QFT ${\cal F}$ of a fundamental field (or fields) $\psi$ that has some unbroken global symmetry given by a group $G$. We can obtain a path integral representation of the reduced density matrix $\rho$ in a region $W$ in the usual form. It is given by the functional matrix
\be
 \rho(\psi_+,\psi_-)= Z(1)^{-1}\,\int^{
 \psi(W+i 0^+)=\psi_+}_
{ \psi(W-i 0^-)=\psi_-} {\cal D}\psi \, e^{-S[\psi]}\,, 
\ee
with $Z(1)= \int {\cal D}\psi \, e^{-S[\psi]}$ the partition function in the plane without boundary conditions on the two sides of the cut $W$. If we are interested in the ``orbifold'' theory ${\cal O}$ of the operators invariant under the symmetry, we have to project this density matrix into the neutral sector. If $W=\cup_{i=1}^m W_i$ is the union of $m$ disjoint regions this projection has to be done in each connected component independently \cite{Casini:2019kex}. This is done by computing 
\be
 \tilde{\rho}(\psi_+,\psi_-)=|G|^{-m}\sum_{g_1,\cdots \,g_m\in G} \rho(g_1\cdots g_m  \psi_+,\psi_- g_m^{-1}\cdots g^{-1}_1)\,,\label{b2}
\ee
where $g_i$ is a twist operator that implements the symmetry group in the region $W_i$ alone and $|G|$ is the number of elements in the group. 
In this way  
\be
\textrm{tr} ( \tilde{\rho} \, X)=\textrm{tr}\left( \rho \, \,\,|G|^{-m}\sum_{g_1,\cdots \,g_m\in G} g_m^{-1}\cdots g^{-1}_1 X g_1\cdots g_m \right)
\ee
 gives the state on the neutral additive algebra on $W$.\footnote{The additive algebra in a region is the one generated by all the algebras of balls included in the region.}

The replica trick then proceeds as usual by computing $\textrm{tr} \tilde{\rho}^n$ by gluing $n$ replicas of the cut plane along the different cuts in cyclic order. The difference with the usual replica trick is that now there are several different partition functions that are added to obtain $\textrm{tr} \tilde{\rho}^n$ due to the sums in (\ref{b2}). 
We get for the Renyi entropy
\be
S_n^{\cal O}(W)=(1-n)^{-1}\textrm{tr} \, \tilde{\rho}^n=(1-n)^{-1}\, \left(\log  \left(|G|^{-m n}\sum_{g_i^k}  Z_{\{g_i^k\}}(n)\right) -n \, \log Z(1)\right)\,.
\ee
 The last term in the brackets corresponds to the normalization of the density matrix where $Z(1)$ is the partition function of the plane without cuts. For $n=1$ the trace eliminates the insertion of group elements and the average is trivial.

These sums are written in terms of group twists operators $\tau_{g_i^k}$, where $i=1,\cdots,m$ denotes a connected component and $k=1,\cdots, n$ is the copy of the plane. 
Due to the cyclic gluing of the copies the partition function depends on the products $\tilde{g}_i^1=g_i^1 (g_i^2)^{-1}, \cdots , \tilde{g}_i^n=g_i^n (g_i^1)^{-1}$  for each connected component $i$. The product of these group elements is the identity,\footnote{While this is not the case of the corresponding twist operators that act on different copies of the space.} 
\be
\tilde{g}_i^1\cdots \tilde{g}_i^n=1\,,\label{producto}
\ee
 and hence there are only $m (n-1)$ independent sums. Another simplification follows from the invariance of the theory under the symmetry group.  This is the freedom of changing variables $\psi\rightarrow g\, \psi$ in each copy. This can be used to eliminate $n$ sums, imposing, for example, that there are no group transformations in one of the connected components and leaving $(m-1) (n-1)$ independent sums over the group elements.\footnote{According to this counting it may then seem that for a single connected component $m=1$ the Renyi entropies of the symmetrized model ${\cal O}$ should coincide with the ones of the full model ${\cal F}$. However, this is a regularization dependent statement. In a lattice, one can see the entropies do not coincide if the algebra of the region is chosen such that the corresponding invariant algebra does not have the same trace dimension \cite{Casini:2019kex}. In a regularization imposed directly in the continuum, such as the one proposed in \cite{Cardy:2016fqc}, where small holes are cut off from the manifold around the boundary of the region and conformal boundary conditions are imposed, the equality will depend on the boundary states at this holes to be invariant under the symmetry.} 

To avoid undefined quantities we compute the mutual information for nearly complementary regions $A$ and $B$. Boundary issues are automatically eliminated. 
If $A,B$ are single component, we have for the Renyi mutual information
\bea
&& I_n^{\cal O}(A,B) = S_{n,\delta}^{\cal O}(A)+S^{\cal O}_{n,\delta}(B)-S^{\cal O}_{n,\delta}(AB) \label{jj}\\
&&= S^{\cal O}_{n,\delta}(A)+S^{\cal O}_{n,\delta}(B)- (1-n)^{-1} \left[\log \left(|G|^{-(n-1)}\sum_{ \{\tilde{g}_A^{k}\}}  Z_{\{\tilde{g}_A^k\}}(n)\right)-n \, \log Z(1) \right]\,,\nonumber 
\eea
where the entropies are computed with a cutoff $\delta$,  we have chosen to keep the group transformations only for the region $A$, and the group elements satisfy the constraint (\ref{producto}).

The partition function $Z(n)$ (for the region $AB$) without group twist insertions  is the expectation value of two replica twist operators $\tau^n_{ A}, (\tau^{n}_{ B})^\dagger $ seated at the boundaries of $A$ and $B$. When these boundaries are near to each other we have an OPE that is dominated by the identity
\be  
\tau^n_{ A} (\tau^{n}_{ B})^\dagger\sim Z(n)+\cdots= e^{-(n-1)\, \left( c_0 \frac{{\cal A}_A+{\cal A}_B}{\delta^{d-2}}-\kappa \frac{{\cal A}}{\epsilon^{d-2}}+\cdots\right)}+\cdots\,,
\ee
where $\delta$ is a cutoff and $\epsilon$ the separation of the boundaries. This gives the area law (and subleading terms) for Renyi mutual  information in the model ${\cal F}$.

The group elements $\tilde{g}_A^k$ in the boundary conditions for the partition function for the different copies can be implemented as the insertion of an additional operator $\prod_k \tau_{\tilde{g}_A^k}$ in the vacuum expectation value in the replicated model. These group twists are of cutoff smearing size $\delta$. 
 The OPE of the full twist operator should give
\be 
 \tau^n_{ A} \prod_{k=1}^{n}  \tau_{\tilde{g}_A^{k}}\,\,(\tau^{n}_{ B})^\dagger 
  \sim Z(n)\, \prod_{k=1}^n \tau_{\tilde{g}^{k}}^\epsilon+\cdots\,,
 \ee
 where $\tau^\epsilon_g$ is some group twist operator over the region seated on the shell with smearing size $\epsilon$. This is because inside $A$, on each copy, the group operation is equivalent to $ \tilde{g}_A^{k}$ and to the identity in $B$; the new twist also obey group rules, and for the identity element $\tilde{g}_A^{k}=1$ we obtain the OPE of the Renyi twist operators.

 Therefore, we get in the limit of small $\epsilon$
\be
|G|^{-(n-1)}\sum_{ \{\tilde{g}_A^{k}\}} Z_{ \{\tilde{g}_A^{k}\}}(n)\sim 
 Z(n)\, \, |G|^{-(n-1)}\sum_{ \{\tilde{g}_A^{k}\}} \langle \prod_{k=1}^n \tau_{\tilde{g}_A^{k}}^\epsilon\rangle \,. 
\ee
Replacing this into (\ref{jj}) we get the leading correction to the Renyi mutual information for small $\epsilon$
\be
 I_n^{\cal O}(A,B) =I_n^{\cal F}(A,B) + \log \left(|G|^{-1} \sum_g \langle \tau^{\epsilon}_g \rangle \right) \,.\label{b10}
\ee
Therefore, for the entropies of $S_n^{\cal O}(A)$ regularized with the mutual information we have the usual replica trick calculation corrected by half this quantity,
\be
S_n^{\cal O}(A)= S_n^{\cal F}(A)+\frac{1}{2} \log \left(|G|^{-1} \sum_g \langle \tau^{\epsilon}_g \rangle \right) \,.
\ee

On each copy, the expectation value of the group of sharp twists 
\be
 \langle  \tau_{g}^\epsilon\rangle\sim \delta_{g,1}+ e^{-c \frac{A}{\epsilon^{d-2}}+\cdots }\,,
\ee
 where only the identity has expectation value that is not suppressed exponentially. 
Then we get
\be
 S_n^{\cal O}(A)  =S_n^{\cal F}(A) -\frac{1}{2}\log |G|\,.
\ee
Note the Renyi mutual informations difference is independent of $n$ in this coincidence limit.\footnote{The mutual information difference is in fact a particular relative entropy for any disjoint $A$ and $B$ \cite{Casini:2019kex}.} A similar behavior (called flat spectrum) has been found in other contexts, for example the boundary entropy \cite{Cardy:2016fqc}, and in holography \cite{Dong:2018seb}.

For a $U(1)$ symmetry an analogous calculation can be done where the averaging is replaced by an integration over the group. If we call $\theta\in (-\pi,\pi)$ to the group parameter ($\theta=0$ corresponds to the identity), we get
\be
  S_n^{\cal O}(A) =S_n^{\cal F}(A) + \frac{1}{2}\log \left((2\pi)^{-1} \int d\theta\,  \langle \tau^{\epsilon}_\theta \rangle \right)\,. 
\ee
Considering that the sharp twists have a Gaussian expectation value \cite{Casini:2019kex}
\be
\langle \tau^{\epsilon}_\theta \rangle\sim e^{-c \, \theta^2 \frac{{\cal A}}{\epsilon^{d-2}}}  \label{gene}
\ee
we get to leading order
\be
 I_n^{\cal O}(A,B) =I_n^{\cal F}(A,B)-\frac{1}{2}\log \frac{{\cal A}}{\epsilon^{d-2}}\,.
\ee
This corrects the logarithmic coefficient in any dimensions by $\left(-\frac{d-2}{2}\right)$ in the mutual information, and half of it for the regularized entropy. The non Abelian case is analogous and the result has an additional factor given by the dimension of the Lie algebra \cite{Casini:2019kex}.\footnote{This follows from the generalization of the Gaussian expectation values (\ref{gene}) to twists operators near the identity in the general Lie group. Interestingly, for a $U(1)$ group the formula (\ref{b10}) of the correction agrees with the entropy in the algebra of group twists but this is not the case for non Abelian groups where there is an additional correction to the entropy \cite{Casini:2019kex}.}  

We make a few remarks. We can think in terms of an effective density matrix description with modular Hamiltonian $H$ and a thermal interpretation of the entropy for this modular energy. Call the thermal partition function ${\cal Z}(n)=\textrm{tr} e^{-n H}$. We have the identification of the Renyi entropies
\be
S_n =(1-n)^{-1} (\log {\cal Z}(n)-n {\cal Z}(1))\,.
\ee
Since we have an effective difference
\be
S_n^{\cal F}-S_n^{\cal O}=-\frac{1}{2}\log \int dg\, \langle \tau_g\rangle\,,
\ee
independent of $n$, the difference is assimilated to a constant term in the free energy
\be
\log {\cal Z}^{\cal F}(n)-\log {\cal Z}^{\cal O}(n)=-\frac{1}{2}\log \int dg\, \langle \tau_g\rangle\,.
\ee
This can be interpreted as the partition function of a decoupled system which will not contribute to the expectation value of the energy and will contribute to the zero temperature entropy of the system. On the other hand, in this effective description, as the statistics of this decoupled system does not depend on the temperature, it would completely degenerate. Note however that this decoupling interpretation needs the limit of small $\epsilon$ and then, in a sense, is also a high-temperature effect on the boundary, which we could interpret as an additional degeneracy of the system of the boundary that it is always in the limit of infinite temperature. 

In this sense, the effect has some similarity to the constant contributions of boundary entropy due to boundary conditions in a CFT. Here there is no change between the models ${\cal F}$ and ${\cal O}$ in the correlation functions of neutral operators inside the region because there is an average over group twisted boundary conditions.   
    
Another interpretation follows by thinking the system $B$ as a purification of the system $A$. Then, the difference in models is because  
 charge fluctuations in $A$ and $B$ compensate each other since the global state is charge neutral, but the entropy in ${\cal O}$ does not take into account the entropy in the fluctuations of charged operators. In this sense, the difference is between the entropies of a density matrix $\rho \sim e^{-H}$ in ${\cal O}$ where we are in the microcanonical ensemble with respect to the charges (not energies), while the charges are allowed to fluctuate freely (with expectation value zero) in ${\cal F}$, a canonical ensemble. Similar effects were studied in BH partition functions (see for example \cite{Sen:2012dw}). In the usual thermodynamical limit, the difference of ensembles is a vanishing small effect that is usually neglected, but for the vacuum EE this difference can be important.      
 
As a final observation, let us consider the case where ${\cal F}$ is a CFT and the group is $U(1)$. For $d=4$ the logarithmic coefficient in the entropy for ${\cal O}$ in a sphere will differ from the anomaly by $-1/2$. We want to elaborate on the failure of the usual proof of the matching of the logarithmic term with the anomaly by mapping the sphere to de Sitter space.  

The orbifold theory ${\cal O}$ will also be a CFT, with the same correlation functions but where only the neutral operators are retained. The stress tensor in both theories is the same operator, and, in even dimensions, the anomaly will be the same. In particular, the expectation value in a conformally flat euclidean space will be
\be
\langle T^\mu_{\mu}(x)\rangle = -2 (-)^{d/2}\, A\, E(x)\,,\label{anon} 
\ee
with $E(x)$ the Euler density (which integrated gives the Euler characteristic of the manifold). Hence, the anomaly $A$ will be the same in both models. The same conclusion can be reached using the definition of the $A$ anomaly in terms of three-point functions of the stress tensor \cite{Erdmenger:1996yc}. 

Therefore, we have a situation where two models have the same $A$ anomaly coefficient and different logarithmic terms in the entropy of a sphere. 
The usual calculation of the logarithmic term by mapping to de Sitter space \cite{Casini:2011kv} depends only on the anomaly though a partition function in a sphere $S^d$, and will erroneously give the same answer to both models
\be
S_{\cal F}=\cdots + (-1)^{\frac{d}{2}-1}\, 4\, A \, \log (R/\epsilon)\,. 
\ee
For the theory ${\cal F}$ this is the correct result, but this is not the case for ${\cal O}$. The mutual information picks up a new term represented in the replica partition function as an average of expectation values of twist operators.  
The new term with the twist expectation values can be thought of as an insertion at the boundary of the region which will be mapped to insertions at the horizon in de Sitter space, with the same results. Then, the partition function has an average over defects (sharp, unitary) on a $S^{d-2}$ surface, and is not the smooth partition function of the fields in $S^{d}$.  

The off-shell computation of the entropy \cite{Dowker:2010bu} follows the thermodynamical formula
\be
S=\int_0^{(2\pi R)^{-1}} dT\, \frac{dE}{dT}\,, \label{ghj}
\ee
by computing the expectation values of the energy density in de Sitter space for different deficit angle. These energy expectation values are the same for the two models and formula (\ref{ghj}) is not able to distinguish between them. The reason is that there is a zero temperature contribution that has to be added to $S^{\cal O}$ that is not contained in this formula which assumes zero entropy for zero temperature.

\section*{Acknowledgments}
We thank discussions with Gonzalo Torroba and Cesar Fosco.  
This work was partially supported by CONICET, CNEA
and Universidad Nacional de Cuyo, Argentina. The work of H. C. and J. M. is partially supported by an It From Qubit grant by the Simons foundation. 
\bibliography{EE}{}

\providecommand{\href}[2]{#2}\begingroup\raggedright\begin{thebibliography}{10}

\bibitem{Solodukhin:2008dh}
S.~N. Solodukhin, ``{Entanglement entropy, conformal invariance and extrinsic
  geometry},'' \href{http://dx.doi.org/10.1016/j.physletb.2008.05.071}{{\em
  Phys. Lett.} {\bfseries B665} (2008) 305--309},
\href{http://arxiv.org/abs/0802.3117}{{\ttfamily arXiv:0802.3117 [hep-th]}}.

\bibitem{Casini:2011kv}
H.~Casini, M.~Huerta, and R.~C. Myers, ``{Towards a derivation of holographic
  entanglement entropy},''
  \href{http://dx.doi.org/10.1007/JHEP05(2011)036}{{\em JHEP} {\bfseries 05}
  (2011) 036},
\href{http://arxiv.org/abs/1102.0440}{{\ttfamily arXiv:1102.0440 [hep-th]}}.

\bibitem{Herzog:2015ioa}
C.~P. Herzog, K.-W. Huang, and K.~Jensen, ``{Universal Entanglement and
  Boundary Geometry in Conformal Field Theory},''
  \href{http://dx.doi.org/10.1007/JHEP01(2016)162}{{\em JHEP} {\bfseries 01}
  (2016) 162},
\href{http://arxiv.org/abs/1510.00021}{{\ttfamily arXiv:1510.00021 [hep-th]}}.

\bibitem{Casini:2010kt}
H.~Casini and M.~Huerta, ``{Entanglement entropy for the n-sphere},''
  \href{http://dx.doi.org/10.1016/j.physletb.2010.09.054}{{\em Phys. Lett.}
  {\bfseries B694} (2011) 167--171},
\href{http://arxiv.org/abs/1007.1813}{{\ttfamily arXiv:1007.1813 [hep-th]}}.

\bibitem{Dowker:2010bu}
J.~S. Dowker, ``{Entanglement entropy for even spheres},''
\href{http://arxiv.org/abs/1009.3854}{{\ttfamily arXiv:1009.3854 [hep-th]}}.

\bibitem{Lohmayer:2009sq}
R.~Lohmayer, H.~Neuberger, A.~Schwimmer, and S.~Theisen, ``{Numerical
  determination of entanglement entropy for a sphere},''
  \href{http://dx.doi.org/10.1016/j.physletb.2010.01.053}{{\em Phys. Lett.}
  {\bfseries B685} (2010) 222--227},
\href{http://arxiv.org/abs/0911.4283}{{\ttfamily arXiv:0911.4283 [hep-lat]}}.

\bibitem{Myers:2010xs}
R.~C. Myers and A.~Sinha, ``{Seeing a c-theorem with holography},''
  \href{http://dx.doi.org/10.1103/PhysRevD.82.046006}{{\em Phys. Rev.}
  {\bfseries D82} (2010) 046006},
\href{http://arxiv.org/abs/1006.1263}{{\ttfamily arXiv:1006.1263 [hep-th]}}.

\bibitem{Eling:2013aqa}
C.~Eling, Y.~Oz, and S.~Theisen, ``{Entanglement and Thermal Entropy of Gauge
  Fields},'' \href{http://dx.doi.org/10.1007/JHEP11(2013)019}{{\em JHEP}
  {\bfseries 11} (2013) 019},
\href{http://arxiv.org/abs/1308.4964}{{\ttfamily arXiv:1308.4964 [hep-th]}}.

\bibitem{Casini:2015dsg}
H.~Casini and M.~Huerta, ``{Entanglement entropy of a Maxwell field on the
  sphere},'' \href{http://dx.doi.org/10.1103/PhysRevD.93.105031}{{\em Phys.
  Rev.} {\bfseries D93} no.~10, (2016) 105031},
\href{http://arxiv.org/abs/1512.06182}{{\ttfamily arXiv:1512.06182 [hep-th]}}.

\bibitem{Casini:2017vbe}
H.~Casini, E.~Test\'e, and G.~Torroba, ``{Markov Property of the Conformal
  Field Theory Vacuum and the a Theorem},''
  \href{http://dx.doi.org/10.1103/PhysRevLett.118.261602}{{\em Phys. Rev.
  Lett.} {\bfseries 118} no.~26, (2017) 261602},
\href{http://arxiv.org/abs/1704.01870}{{\ttfamily arXiv:1704.01870 [hep-th]}}.

\bibitem{Casini:2018kzx}
H.~Casini, E.~Teste, and G.~Torroba, ``{All the entropies on the light-cone},''
  \href{http://dx.doi.org/10.1007/JHEP05(2018)005}{{\em JHEP} {\bfseries 05}
  (2018) 005},
\href{http://arxiv.org/abs/1802.04278}{{\ttfamily arXiv:1802.04278 [hep-th]}}.

\bibitem{Pretko:2018yxl}
M.~Pretko, ``{On the Entanglement Entropy of Maxwell Theory: A Condensed Matter
  Perspective},'' \href{http://dx.doi.org/10.1007/JHEP12(2018)102}{{\em JHEP}
  {\bfseries 12} (2018) 102},
\href{http://arxiv.org/abs/1801.01158}{{\ttfamily arXiv:1801.01158 [hep-th]}}.

\bibitem{Grover:2011fa}
T.~Grover, A.~M. Turner, and A.~Vishwanath, ``{Entanglement Entropy of Gapped
  Phases and Topological Order in Three dimensions},''
  \href{http://dx.doi.org/10.1103/PhysRevB.84.195120}{{\em Phys. Rev.}
  {\bfseries B84} (2011) 195120},
\href{http://arxiv.org/abs/1108.4038}{{\ttfamily arXiv:1108.4038
  [cond-mat.str-el]}}.

\bibitem{Liu:2012eea}
H.~Liu and M.~Mezei, ``{A Refinement of entanglement entropy and the number of
  degrees of freedom},'' \href{http://dx.doi.org/10.1007/JHEP04(2013)162}{{\em
  JHEP} {\bfseries 04} (2013) 162},
\href{http://arxiv.org/abs/1202.2070}{{\ttfamily arXiv:1202.2070 [hep-th]}}.

\bibitem{Casini:2019kex}
H.~Casini, M.~Huerta, J.~M. Mag\'an, and D.~Pontello, ``{Entanglement entropy
  and superselection sectors I. Global symmetries},''
\href{http://arxiv.org/abs/1905.10487}{{\ttfamily arXiv:1905.10487 [hep-th]}}.

\bibitem{Casini:2015woa}
H.~Casini, M.~Huerta, R.~C. Myers, and A.~Yale, ``{Mutual information and the
  F-theorem},'' \href{http://dx.doi.org/10.1007/JHEP10(2015)003}{{\em JHEP}
  {\bfseries 10} (2015) 003},
\href{http://arxiv.org/abs/1506.06195}{{\ttfamily arXiv:1506.06195 [hep-th]}}.

\bibitem{Casini:2007dk}
H.~Casini, ``{Entropy localization and extensivity in the semiclassical black
  hole evaporation},'' \href{http://dx.doi.org/10.1103/PhysRevD.79.024015}{{\em
  Phys. Rev.} {\bfseries D79} (2009) 024015},
\href{http://arxiv.org/abs/0712.0403}{{\ttfamily arXiv:0712.0403 [hep-th]}}.

\bibitem{Itzykson:1980rh}
C.~Itzykson and J.~B. Zuber, {\em {Quantum Field Theory}}.
\newblock International Series In Pure and Applied Physics. McGraw-Hill, New
  York, 1980.
\newblock
\url{http://dx.doi.org/10.1063/1.2916419}.
\newblock

\bibitem{Dorigoni:2009ra}
D.~Dorigoni and V.~S. Rychkov, ``{Scale Invariance + Unitarity => Conformal
  Invariance?},''
\href{http://arxiv.org/abs/0910.1087}{{\ttfamily arXiv:0910.1087 [hep-th]}}.

\bibitem{Donnelly:2015hxa}
W.~Donnelly and A.~C. Wall, ``{Geometric entropy and edge modes of the
  electromagnetic field},''
  \href{http://dx.doi.org/10.1103/PhysRevD.94.104053}{{\em Phys. Rev.}
  {\bfseries D94} no.~10, (2016) 104053},
\href{http://arxiv.org/abs/1506.05792}{{\ttfamily arXiv:1506.05792 [hep-th]}}.

\bibitem{Donnelly:2014fua}
W.~Donnelly and A.~C. Wall, ``{Entanglement entropy of electromagnetic edge
  modes},'' \href{http://dx.doi.org/10.1103/PhysRevLett.114.111603}{{\em Phys.
  Rev. Lett.} {\bfseries 114} no.~11, (2015) 111603},
\href{http://arxiv.org/abs/1412.1895}{{\ttfamily arXiv:1412.1895 [hep-th]}}.

\bibitem{Soni:2016ogt}
R.~M. Soni and S.~P. Trivedi, ``{Entanglement entropy in (3 + 1)-d free U(1)
  gauge theory},'' \href{http://dx.doi.org/10.1007/JHEP02(2017)101}{{\em JHEP}
  {\bfseries 02} (2017) 101},
\href{http://arxiv.org/abs/1608.00353}{{\ttfamily arXiv:1608.00353 [hep-th]}}.

\bibitem{Moitra:2018lxn}
U.~Moitra, R.~M. Soni, and S.~P. Trivedi, ``{Entanglement Entropy, Relative
  Entropy and Duality},'' \href{http://dx.doi.org/10.1007/JHEP08(2019)059}{{\em
  JHEP} {\bfseries 08} (2019) 059},
\href{http://arxiv.org/abs/1811.06986}{{\ttfamily arXiv:1811.06986 [hep-th]}}.

\bibitem{Huang:2014pfa}
K.-W. Huang, ``{Central Charge and Entangled Gauge Fields},''
  \href{http://dx.doi.org/10.1103/PhysRevD.92.025010}{{\em Phys. Rev.}
  {\bfseries D92} no.~2, (2015) 025010},
\href{http://arxiv.org/abs/1412.2730}{{\ttfamily arXiv:1412.2730 [hep-th]}}.

\bibitem{Srednicki:1993im}
M.~Srednicki, ``{Entropy and area},''
  \href{http://dx.doi.org/10.1103/PhysRevLett.71.666}{{\em Phys. Rev. Lett.}
  {\bfseries 71} (1993) 666--669},
\href{http://arxiv.org/abs/hep-th/9303048}{{\ttfamily arXiv:hep-th/9303048
  [hep-th]}}.

\bibitem{Casini:2013rba}
H.~Casini, M.~Huerta, and J.~A. Rosabal, ``{Remarks on entanglement entropy for
  gauge fields},'' \href{http://dx.doi.org/10.1103/PhysRevD.89.085012}{{\em
  Phys. Rev.} {\bfseries D89} no.~8, (2014) 085012},
\href{http://arxiv.org/abs/1312.1183}{{\ttfamily arXiv:1312.1183 [hep-th]}}.

\bibitem{Zuo:2016knh}
F.~Zuo, ``{A note on electromagnetic edge modes},''
\href{http://arxiv.org/abs/1601.06910}{{\ttfamily arXiv:1601.06910 [hep-th]}}.

\bibitem{Buividovich:2008gq}
P.~V. Buividovich and M.~I. Polikarpov, ``{Entanglement entropy in gauge
  theories and the holographic principle for electric strings},''
  \href{http://dx.doi.org/10.1016/j.physletb.2008.10.032}{{\em Phys. Lett.}
  {\bfseries B670} (2008) 141--145},
\href{http://arxiv.org/abs/0806.3376}{{\ttfamily arXiv:0806.3376 [hep-th]}}.

\bibitem{Donnelly:2008vx}
W.~Donnelly, ``{Entanglement entropy in loop quantum gravity},''
  \href{http://dx.doi.org/10.1103/PhysRevD.77.104006}{{\em Phys. Rev.}
  {\bfseries D77} (2008) 104006},
\href{http://arxiv.org/abs/0802.0880}{{\ttfamily arXiv:0802.0880 [gr-qc]}}.

\bibitem{Donnelly:2011hn}
W.~Donnelly, ``{Decomposition of entanglement entropy in lattice gauge
  theory},'' \href{http://dx.doi.org/10.1103/PhysRevD.85.085004}{{\em Phys.
  Rev.} {\bfseries D85} (2012) 085004},
\href{http://arxiv.org/abs/1109.0036}{{\ttfamily arXiv:1109.0036 [hep-th]}}.

\bibitem{Ghosh:2015iwa}
S.~Ghosh, R.~M. Soni, and S.~P. Trivedi, ``{On The Entanglement Entropy For
  Gauge Theories},'' \href{http://dx.doi.org/10.1007/JHEP09(2015)069}{{\em
  JHEP} {\bfseries 09} (2015) 069},
\href{http://arxiv.org/abs/1501.02593}{{\ttfamily arXiv:1501.02593 [hep-th]}}.

\bibitem{Soni:2015yga}
R.~M. Soni and S.~P. Trivedi, ``{Aspects of Entanglement Entropy for Gauge
  Theories},'' \href{http://dx.doi.org/10.1007/JHEP01(2016)136}{{\em JHEP}
  {\bfseries 01} (2016) 136},
\href{http://arxiv.org/abs/1510.07455}{{\ttfamily arXiv:1510.07455 [hep-th]}}.

\bibitem{VanAcoleyen:2015ccp}
K.~Van~Acoleyen, N.~Bultinck, J.~Haegeman, M.~Marien, V.~B. Scholz, and
  F.~Verstraete, ``{The entanglement of distillation for gauge theories},''
  \href{http://dx.doi.org/10.1103/PhysRevLett.117.131602}{{\em Phys. Rev.
  Lett.} {\bfseries 117} no.~13, (2016) 131602},
\href{http://arxiv.org/abs/1511.04369}{{\ttfamily arXiv:1511.04369
  [quant-ph]}}.

\bibitem{Donnelly:2014gva}
W.~Donnelly, ``{Entanglement entropy and nonabelian gauge symmetry},''
  \href{http://dx.doi.org/10.1088/0264-9381/31/21/214003}{{\em Class. Quant.
  Grav.} {\bfseries 31} no.~21, (2014) 214003},
\href{http://arxiv.org/abs/1406.7304}{{\ttfamily arXiv:1406.7304 [hep-th]}}.

\bibitem{Aoki:2015bsa}
S.~Aoki, T.~Iritani, M.~Nozaki, T.~Numasawa, N.~Shiba, and H.~Tasaki, ``{On the
  definition of entanglement entropy in lattice gauge theories},''
  \href{http://dx.doi.org/10.1007/JHEP06(2015)187}{{\em JHEP} {\bfseries 06}
  (2015) 187},
\href{http://arxiv.org/abs/1502.04267}{{\ttfamily arXiv:1502.04267 [hep-th]}}.

\bibitem{Radicevic:2015sza}
D.~Radicevic, ``{Entanglement in Weakly Coupled Lattice Gauge Theories},''
  \href{http://dx.doi.org/10.1007/JHEP04(2016)163}{{\em JHEP} {\bfseries 04}
  (2016) 163},
\href{http://arxiv.org/abs/1509.08478}{{\ttfamily arXiv:1509.08478 [hep-th]}}.

\bibitem{Hung:2015fla}
L.-Y. Hung and Y.~Wan, ``{Revisiting Entanglement Entropy of Lattice Gauge
  Theories},'' \href{http://dx.doi.org/10.1007/JHEP04(2015)122}{{\em JHEP}
  {\bfseries 04} (2015) 122},
\href{http://arxiv.org/abs/1501.04389}{{\ttfamily arXiv:1501.04389 [hep-th]}}.

\bibitem{Radicevic:2016tlt}
D.~Radicevic, ``{Entanglement Entropy and Duality},''
  \href{http://dx.doi.org/10.1007/JHEP11(2016)130}{{\em JHEP} {\bfseries 11}
  (2016) 130},
\href{http://arxiv.org/abs/1605.09396}{{\ttfamily arXiv:1605.09396 [hep-th]}}.

\bibitem{Donnelly:2016mlc}
W.~Donnelly, B.~Michel, and A.~Wall, ``{Electromagnetic Duality and
  Entanglement Anomalies},''
  \href{http://dx.doi.org/10.1103/PhysRevD.96.045008}{{\em Phys. Rev.}
  {\bfseries D96} no.~4, (2017) 045008},
\href{http://arxiv.org/abs/1611.05920}{{\ttfamily arXiv:1611.05920 [hep-th]}}.

\bibitem{Donnelly:2012st}
W.~Donnelly and A.~C. Wall, ``{Do gauge fields really contribute negatively to
  black hole entropy?},''
  \href{http://dx.doi.org/10.1103/PhysRevD.86.064042}{{\em Phys. Rev.}
  {\bfseries D86} (2012) 064042},
\href{http://arxiv.org/abs/1206.5831}{{\ttfamily arXiv:1206.5831 [hep-th]}}.

\bibitem{Kabat:1995eq}
D.~N. Kabat, ``{Black hole entropy and entropy of entanglement},''
  \href{http://dx.doi.org/10.1016/0550-3213(95)00443-V}{{\em Nucl. Phys.}
  {\bfseries B453} (1995) 281--299},
\href{http://arxiv.org/abs/hep-th/9503016}{{\ttfamily arXiv:hep-th/9503016
  [hep-th]}}.

\bibitem{Solodukhin:2012jh}
S.~N. Solodukhin, ``{Remarks on effective action and entanglement entropy of
  Maxwell field in generic gauge},''
  \href{http://dx.doi.org/10.1007/JHEP12(2012)036}{{\em JHEP} {\bfseries 12}
  (2012) 036},
\href{http://arxiv.org/abs/1209.2677}{{\ttfamily arXiv:1209.2677 [hep-th]}}.

\bibitem{Lin:2018bud}
J.~Lin and D.~Radicevic, ``{Comments on Defining Entanglement Entropy},''
\href{http://arxiv.org/abs/1808.05939}{{\ttfamily arXiv:1808.05939 [hep-th]}}.

\bibitem{Lin:2017uzr}
J.~Lin, ``{Ryu-Takayanagi Area as an Entanglement Edge Term},''
\href{http://arxiv.org/abs/1704.07763}{{\ttfamily arXiv:1704.07763 [hep-th]}}.

\bibitem{Donnelly:2016auv}
W.~Donnelly and L.~Freidel, ``{Local subsystems in gauge theory and gravity},''
  \href{http://dx.doi.org/10.1007/JHEP09(2016)102}{{\em JHEP} {\bfseries 09}
  (2016) 102},
\href{http://arxiv.org/abs/1601.04744}{{\ttfamily arXiv:1601.04744 [hep-th]}}.

\bibitem{Cappelli:2000fe}
A.~Cappelli and G.~D'Appollonio, ``{On the trace anomaly as a measure of
  degrees of freedom},''
  \href{http://dx.doi.org/10.1016/S0370-2693(00)00809-1}{{\em Phys. Lett.}
  {\bfseries B487} (2000) 87--95},
\href{http://arxiv.org/abs/hep-th/0005115}{{\ttfamily arXiv:hep-th/0005115
  [hep-th]}}.

\bibitem{DeNardo:1996kp}
L.~De~Nardo, D.~V. Fursaev, and G.~Miele, ``{Heat kernel coefficients and
  spectra of the vector Laplacians on spherical domains with conical
  singularities},'' \href{http://dx.doi.org/10.1088/0264-9381/14/5/013}{{\em
  Class. Quant. Grav.} {\bfseries 14} (1997) 1059--1078},
\href{http://arxiv.org/abs/hep-th/9610011}{{\ttfamily arXiv:hep-th/9610011
  [hep-th]}}.

\bibitem{Fursaev:1996uz}
D.~V. Fursaev and G.~Miele, ``{Cones, spins and heat kernels},''
  \href{http://dx.doi.org/10.1016/S0550-3213(96)00631-1}{{\em Nucl. Phys.}
  {\bfseries B484} (1997) 697--723},
\href{http://arxiv.org/abs/hep-th/9605153}{{\ttfamily arXiv:hep-th/9605153
  [hep-th]}}.

\bibitem{Calabrese:2004eu}
P.~Calabrese and J.~L. Cardy, ``{Entanglement entropy and quantum field
  theory},'' \href{http://dx.doi.org/10.1088/1742-5468/2004/06/P06002}{{\em J.
  Stat. Mech.} {\bfseries 0406} (2004) P06002},
\href{http://arxiv.org/abs/hep-th/0405152}{{\ttfamily arXiv:hep-th/0405152
  [hep-th]}}.

\bibitem{Headrick:2010zt}
M.~Headrick, ``{Entanglement Renyi entropies in holographic theories},''
  \href{http://dx.doi.org/10.1103/PhysRevD.82.126010}{{\em Phys. Rev.}
  {\bfseries D82} (2010) 126010},
\href{http://arxiv.org/abs/1006.0047}{{\ttfamily arXiv:1006.0047 [hep-th]}}.

\bibitem{Cardy.esferaslejanas}
J.~Cardy, ``Some results on the mutual information of disjoint regions in
  higher dimensions,'' {\em Journal of Physics A: Mathematical and Theoretical}
  {\bfseries 46} no.~28, (2013) 285402.
  \url{http://stacks.iop.org/1751-8121/46/i=28/a=285402}.

\bibitem{Bousso:2014uxa}
R.~Bousso, H.~Casini, Z.~Fisher, and J.~Maldacena, ``{Entropy on a null surface
  for interacting quantum field theories and the Bousso bound},''
  \href{http://dx.doi.org/10.1103/PhysRevD.91.084030}{{\em Phys. Rev.}
  {\bfseries D91} no.~8, (2015) 084030},
\href{http://arxiv.org/abs/1406.4545}{{\ttfamily arXiv:1406.4545 [hep-th]}}.

\bibitem{Casini:2017roe}
H.~Casini, E.~Teste, and G.~Torroba, ``{Modular Hamiltonians on the null plane
  and the Markov property of the vacuum state},''
  \href{http://dx.doi.org/10.1088/1751-8121/aa7eaa}{{\em J. Phys.} {\bfseries
  A50} no.~36, (2017) 364001},
\href{http://arxiv.org/abs/1703.10656}{{\ttfamily arXiv:1703.10656 [hep-th]}}.

\bibitem{Hollands:2017dov}
S.~Hollands and K.~Sanders, ``{Entanglement measures and their properties in
  quantum field theory},''
\href{http://arxiv.org/abs/1702.04924}{{\ttfamily arXiv:1702.04924
  [quant-ph]}}.

\bibitem{Cardy:2016fqc}
J.~Cardy and E.~Tonni, ``{Entanglement hamiltonians in two-dimensional
  conformal field theory},''
  \href{http://dx.doi.org/10.1088/1742-5468/2016/12/123103}{{\em J. Stat.
  Mech.} {\bfseries 1612} no.~12, (2016) 123103},
\href{http://arxiv.org/abs/1608.01283}{{\ttfamily arXiv:1608.01283
  [cond-mat.stat-mech]}}.

\bibitem{Dong:2018seb}
X.~Dong, D.~Harlow, and D.~Marolf, ``{Flat entanglement spectra in fixed-area
  states of quantum gravity},''
\href{http://arxiv.org/abs/1811.05382}{{\ttfamily arXiv:1811.05382 [hep-th]}}.

\bibitem{Sen:2012dw}
A.~Sen, ``{Logarithmic Corrections to Schwarzschild and Other Non-extremal
  Black Hole Entropy in Different Dimensions},''
  \href{http://dx.doi.org/10.1007/JHEP04(2013)156}{{\em JHEP} {\bfseries 04}
  (2013) 156},
\href{http://arxiv.org/abs/1205.0971}{{\ttfamily arXiv:1205.0971 [hep-th]}}.

\bibitem{Erdmenger:1996yc}
J.~Erdmenger and H.~Osborn, ``{Conserved currents and the energy momentum
  tensor in conformally invariant theories for general dimensions},''
  \href{http://dx.doi.org/10.1016/S0550-3213(96)00545-7}{{\em Nucl. Phys.}
  {\bfseries B483} (1997) 431--474},
\href{http://arxiv.org/abs/hep-th/9605009}{{\ttfamily arXiv:hep-th/9605009
  [hep-th]}}.

\end{thebibliography}\endgroup
\bibliographystyle{utphys}

\end{document}